\documentclass[aps,prd,superscriptaddress,reprint,floatfix,nofootinbib]{revtex4-2}
\usepackage{fancyhdr}
\usepackage{graphicx}  
\usepackage{dcolumn}   
\usepackage{bm}        
\usepackage{amssymb}   
\usepackage{amsmath}
\usepackage{epsfig}
\usepackage{rotating}
\usepackage{slashed}
\usepackage{float}
\usepackage[normalem]{ulem} 
\usepackage{fancybox}
\usepackage[caption=false]{subfig}
\newcommand{\pr}{\prime}
\newcommand{\ppr}{{\prime\prime}}
\usepackage[usenames, dvipsnames]{color}
\usepackage[colorlinks=true,citecolor=Green,linkcolor=Blue,urlcolor=Black]{hyperref}
\newcommand{\nn}{\nonumber}
\newcommand\ringring[1]{%
	{
		\mathop{\kern0pt #1}\limits^{
			\vbox to-1.85ex{
				\kern-2ex 
				\hbox to 0pt{\hss\normalfont\kern.1em \r{}\kern-.45em \r{}\hss}%
				\vss 
			}
		}
	}
}
\makeatletter
\renewcommand{\p@subsection}{}
\renewcommand{\p@subsubsection}{}
\def\l@subsubsection#1#2{}
\makeatother
\DeclareMathOperator\artanh{artanh}

\usepackage{cleveref}
\begin{document}
	
	
	\title{Unruh-DeWitt detectors in cosmological spacetimes.}
	
	\author{Aindri\'u Conroy}
	\email{aindriu.conroy@matfyz.cuni.cz}
\affiliation{Institute of Theoretical Physics, Faculty of Mathematics and Physics,
	Charles University, V Hole\v{s}ovi\v{c}k\'ach 2, Prague, 18000, Czech Republic}
	\affiliation{Centre for Astrophysics and Relativity,
		Dublin City University,
		Glasnevin,
		Dublin 9, Rep. of Ireland.
	}
		\date{\today}
	
	\begin{abstract}
		We analyse the response and thermal behaviour of an Unruh-DeWitt detector as it travels through cosmological spacetimes, with special reference to the question of how to define surface gravity and temperature in dynamical spacetimes. Working within the quantum field theory on curved spacetime approximation, we consider a detector as it travels along geodesic and accelerated Kodama trajectories in de Sitter and asymptotically de Sitter FLRW spacetimes. By modelling the temperature of the detector using the detailed-balance form of the Kubo–Martin–Schwinger (KMS) conditions as it thermalises, we can better understand the thermal behaviour of the detector as it interacts with the quantum field, and use this to compare competing definitions of surface gravity and temperature that persist in the literature. These include the approaches of Hayward-Kodama, Ashtekar et al., Fodor et al., and Nielsen-Visser. While these are most often examined within the context of a dynamical black hole, here we shift  focus to surface gravity on the evolving cosmological horizon.
	\end{abstract}
	
	\maketitle
	\section{Introduction}
	We understand surface gravity, in the Newtonian sense, as the acceleration due to the force of gravity. This is the acceleration characterised by the familiar $g \sim m/r^2$, ubiquitous in Newtonian mechanics. In the wider, astronomical sense, we think of surface gravity as the acceleration required to keep a point particle in place on a given surface where the mass is taken to be negligible and the surface we have in mind is, for instance, the surface of a celestial body. This concept is muddied somewhat in the context of a relativistic black hole. Firstly, in place of the common-sense notion of the surface of some astronomical body, we have the abstract concept of an event horizon. This horizon is generated by the failure of null geodesics to reach infinity and we can think of this surface as the causal boundary which obscures information to a distant observer. Crucially, however, the acceleration measured on this horizon blows up as the radius $r$ approaches zero.
	
	Enter the concept of a Killing vector. In stationary spacetimes, such as a static black hole, we interpret the event horizon as the two-dimensional null hypersurface formed where null and timelike Killing vectors coincide. This in itself does not resolve the issue of the four-acceleration diverging on the horizon. Instead, we amend our interpretation through the red-shift factor $V$, where now the surface gravity is understood via $\kappa \sim V\times A$ where $A$ is the magnitude of the four-acceleration. The red-shift factor serves to shift the locally-applied force (acceleration) required to keep a point particle in place at some radius $r$ to infinity through, what is often referred to as, an `infinitely-long massless string'. 
	This interpretation ensures that the surface gravity remains regular on the horizon and leads to a consistent notion of temperature $T=\kappa/2\pi$ on the event horizon of a black hole, the so-called \emph{Hawking temperature}. This, in a nutshell, is the classical notion of surface-gravity in \emph{General Relativity}, defined as it is through the geometric properties of the spacetime.

	As Killing vectors persist only in stationary spacetimes, what then of dynamical spacetimes such as a black hole with an evolving horizon or an expanding universe? Without a well-defined notion of a Killing horizon, the framework for understanding physical quantities such as four-acceleration, surface gravity, and temperature collapses. It is evident then that we require some alternative prescription for surface gravity in dynamical spacetimes.
	
Several propositions have been put forward in the context of dynamical black holes to better understand the nature of temperature on an horizon which evolves with time. First among these is the approach of Kodama and Hayward. This approach rests upon the insight of Hideo Kodama in Ref.~\cite{Kodama} of constructing a divergence-free vector field which mimics the behaviour of a Killing vector in spherically-symmetric, dynamical backgrounds by describing a trapping surface (or apparent horizon) which is suitably analogous to the event horizon of a stationary black hole. 
 This approach was later built upon by Hayward in Ref.~\cite{Hayward:1997jp} to give a geometric description of surface gravity on an evolving horizon which we call \emph{Hayward-Kodama surface gravity}. 

Various other approaches persist in the literature including the effective surface gravity put forward by Ashtekar et al.~in Ref.~\cite{Ashtekar:2004cn} and tuned to give the correct Killing horizon behaviour of a Schwarzschild black hole in the static limit; the Fodor approach to surface gravity which, in the case of a dynamical black hole, relies on the presence of a marginally outer-trapped surface upon which measurements are made; and Nielsen-Visser surface gravity which is formulated through the Misner-Sharp-Hernandez mass $M=M(t,\vec r)$, the dynamical nature of which is a result of the non-trivial mass-energy exchange which occurs in dynamical black holes, see Ref.~\cite{Nielsen:2005af}. 

Here, we investigate all of these proposals but with a shift in focus away from dynamical black holes towards (dynamical) cosmological backgrounds. While each definition is distinct in its a priori formulation, these approaches boil down to two categories when evaluated on the cosmological apparent horizon within the framework of a geometrically-flat FLRW spacetime. In the first instance we have the Hayward-Kodama surface gravity defined via 
\begin{equation}
		\kappa_{HK}\equiv	\frac{1}{\tilde{r}_{AH}}\left[1-\frac{\dot{\tilde{r}}_{AH}}{2H\tilde{r}_{AH}}\right],
\end{equation}
while the alternatives of Ashtekar, Fodor, and Nielsen-Visser all reduce to 
\begin{equation}
	\kappa_{eff}\equiv 	\frac{1}{\tilde{r}_{AH}},
\end{equation}
when evaluated on the cosmological apparent horizon $r_{AH}=H^{-1}$.

	The article aims to analyse the response and thermal behaviour of a particle detector in cosmological spacetimes with special reference to this issue of surface gravity, and its relation to temperature, in dynamical spacetimes. We work within the framework of the Unruh-DeWitt particle model which lends an operational meaning to the notion of a `particle' in Quantum Field Theory in curved spacetime (QFTCS). The particle ambiguity in QFTCS stems from the absence of a uniquely-defined vacuum, Ref.~\cite{birrell1984quantum}, that is to say that there is no natural choice of quantum state which corresponds to the physical absence of `particles'. While a particle may be observed by one detector tuned to a particular vacuum, it will not necessarily be observed by a detector coupled to a field in a different quantum state. This is true even in Minkowski space where the vacuum state is chosen not because it is uniquely defined but because it is the vacuum state which agrees with all measuring devices passing along an inertial trajectory. This provides flat-space field theory with a preferred choice of vacuum unlike its curved space cousin. Thus, in a generic curved spacetime, one can not identify a global time function to distinguish between positive frequency and negative frequency modes leading to an ambiguity in the particle concept. This means that what we call a `particle' in curved space can not be universally understood, meaning that the very notion is ill-defined.

The canonical solution is to treat fields rather than particles as the fundamental object of interest. However, in a seminal 1976 paper, Ref.~\cite{Unruh:1976}, Bill Unruh offered a well-defined operational meaning to the concept of a particle by coupling a quantum field to a two-level idealized atom and considering the absorption and emission of field quanta by the atom. This is the so-called \emph{Unruh-DeWitt detector model} and we work within this framework. 

We have in mind a measuring device which is an idealised quantum mechanical object, travelling through spacetime on a given trajectory. Just as an electron moves from its ground state to its excited state through the absorption of a photon, the device we have in mind is itself a two-level atom where interaction with the quantum field governs the transition from ground to excited state (and vice versa).  We interpret this atomic excitation as the device registering a `particle' and the device itself as a `particle detector', which imbues an \emph{operational} understanding to the notion of a particle. 

If such a detector is uniformly accelerating, it will observe black-body radiation according to the \emph{Unruh effect}, while quantum fluctuations on the event horizon of a black hole will result in the detector registering low-energy radiation apropos the \emph{Hawking effect}. Much work in examining the response and thermal behaviour of a detector has been carried out in the context of black holes, see for instance Refs.~\cite{HodgkinsonLoukoOttewill,Louko2014,LoukoHodgkinson:2011pc,LoukoSatz2006,LoukoSatz2008,LoukoToussaint,Satz2007,MartinezLouko,Brenna2016,Mann2020,Conroy13}. Here, we focus on the expanding universe so that in place of the event horizon we have the dynamical cosmological apparent horizon, and tiny fluctuations on this surface will result in the detector similarly registering low-energy radiation. This is a consequence of the analogous phenomenon of \emph{cosmological particle creation} first introduced by Leonard Parker in Ref.~\cite{Parker1968}.

The article is organised as follows. In Section \ref{sec:theory}, we review the theory behind the Unruh-deWitt particle detector before introducing the competing concepts of surface gravity in dynamical spacetimes in Section \ref{sec:SurfaceGravity}. In Section \ref{sec:Trajectory}, we briefly outline the geodesic and accelerated trajectories to be considered before examining, in Section \ref{sec:dS}, the response and thermal behaviour of a detector passing along these trajectories in the de Sitter universe. In this section, we introduce the methodologies at are disposal in a well-studied spacetime while the results, particular in the context of the proposed anti-Unruh effect, go beyond what is found in the literature \cite{Acquaviva:2011vq,Ali:2020gij}. We then extend our study in Section \ref{sec:FLRW} to a less-restrictive FLRW spacetime with a choice of scale factor that is both tractable in terms of a Kodama detector and asymptotes to de Sitter space. We end with a discussion of our results and future outlook in Section \ref{sec:Discussion}.

\section{Particle detector theory}
\label{sec:theory}
As with Refs.~\cite{HodgkinsonLoukoOttewill,Satz2007}, the particle detector model we are considering has been furnished with a switching function $\chi$ which controls how the interaction with the quantum field is turned on and off. We assume initially that such a function is smooth and has compact support, while the response function which measures these transitions is well-defined provided that the quantum state of the field is Hadamard. While the switching function is initially assumed to be smooth, we will see later that it approaches a sharp-switching step function when an appropriate limit is taken

Suppose then that the particle detector travels along a world line $x^\mu(\tau)$, where $\tau$ is the proper time of the detector. We model the interaction between the detector and the real-valued quantum field $\hat\varphi(x)$ via the interaction Hamiltonian
\begin{equation}
	H_{int}=c\chi(\tau)\hat m(\tau)\hat\varphi(x),
\end{equation}
where $c$ is a coupling constant assumed to be small, so that we can treat the system as a perturbation around the free Hamiltonian;  $\hat m(\tau)$ is the detector's monopole moment operator; and we choose $\hat\varphi(x)$ to be massless. 

Before the detector and the quantum field interact, we suppose that the field $\hat\varphi(x)$ is in some initial Hadamard state $|\Phi\rangle$ on a given background, while the detector is in its ground state $|E_0\rangle$. When interaction takes place, the field $\hat\varphi(x)$ transitions from its initial state $|\Phi\rangle$ to a final state, which we shall call $|\Phi^\prime \rangle$, while the detector undergoes a transition from ground state $|E_0\rangle$ to excited state $|E\rangle$, the specifics of which depend on the trajectory of the detector in the given spacetime.

The probability that the detector undergoes this transition during the interaction time is the quantity of interest. When the detector leaves its ground state, the eigenvalue for the state $|E\rangle$ may be positive or negative. In the case where $E>E_0$, the interpretation is that the detector has absorbed field quanta while for $E<E_0$, the detector has emitted field quanta. To first order in perturbation theory (assuming that $c$ is small), the probability for this transition is
\begin{equation}
	\label{transitionprob}
	P(\omega)= c^2\rvert\langle E|\hat m(0)|E_0\rangle\rvert^2 {\cal F}(\omega),
\end{equation}
where we have defined the \emph{response function}\footnote{Here, we have chosen $u=\tau$ and $s=\tau-\tau^\prime$ in the region where $\tau>\tau^\prime$, while $u=\tau^\prime$ and $s=\tau^\prime-\tau$ in the region $\tau^\prime>\tau$.} 
\begin{align}
	\label{eq:response}
	\mathcal{F}(\omega)&\equiv 2\,\lim_{\epsilon\to 0^{+}} \Re \int_{-\infty}^{\infty}du\,\chi(u)\times \nn\\&\times\int_{0}^{\infty}ds\,\chi(u-s)e^{-i \,\omega\,s}W_\epsilon(u, u-s),
\end{align}
written in terms of  the energy gap $\omega\equiv E-E_0$, where the quantity $W_{\epsilon}(x,x')$ is the Wightman Green function for the massless scalar wave equation.

We note that all of the dependence on the quantum field $\hat{\varphi}(x)$, the trajectory $x^\mu(\tau)$ and the switching function $\chi(\tau)$ is contained in the response function while the factor $ c^2\rvert\langle E|\hat m(0)|E_0\rangle\rvert^2$ in Eq.~\eqref{transitionprob} is straightforward to compute and depends only on the structure of the detector which we have modelled as a monopole. Hence, it is typical in the literature to refer to $\mathcal{F}(\omega)$ as the transition probability. In terms of the Wightman two-point function, we have in mind the distribution\footnote{The spatial trajectories are encoded in $\mathbf{x}=\mathbf{x}(\tau)$ while $\eta=\eta(\tau)$ is the conformal time parameter and $\tau=u$ is proper time throughout. Both spatial and temporal components are contained within the parameter $x=x(\tau)$ while the $^\prime$ notation signifies $x^\prime=x(\tau-s)$, $\eta^\prime=\eta(\tau-s)$, etc. The scale factor is given by $a$.} 
\begin{equation}
	\label{Wightman}
	W_\epsilon(x;x^\pr)	=\frac{1}{(2\pi)^{2}}\frac{1}{a(\eta)a(\eta^{\prime})}\frac{1}{-|\eta-\eta^{\prime}-i\epsilon|^{2}+|\mathbf{x}-\mathbf{x}^{\prime}|^{2}},
\end{equation}
for a field in the \emph{conformal vacuum}\footnote{By `conformal vacuum' we mean a mode decomposition based on the conformally-flat nature of the FLRW metric which reduces to the flat space mode decomposition when $a(t)=1$.}, the details of which can be found in Appendix \ref{sec:Mode}.

The Wightman two-point function is regular except when the arguments are evaluated at the same spacetime point, i.e. when $x=x^\prime$. In this way, it follows the Hadamard singularity structure and requires a regularisation procedure to ensure meaningful results. The regularisation prescription employed in Eq.~\eqref{Wightman} follows what is sometimes referred to as the `Feynman prescription' of inserting a small parameter $i\epsilon$ to ensure that $W_\epsilon(x;x^\pr)$ is a well-defined distribution. This choice of regularisation was queried by Schlicht in Ref. \cite{Schlicht:2003iy} following his observation that such a regularisation gives unphysical results when used to reproduce the thermal Unruh spectrum seen by an accelerated observer. 

The issue stems from the switching function $\chi(\tau)$ and the contradicting requirements of \emph{instantaneous} excitation and the integrand itself, with the former requiring a sharp cut-off in order to take an instantaneous measurement, and the latter required to be smooth. While Schlicht's path was to suggest an alternative regularization, it was shown in Refs. \cite{LoukoHodgkinson:2011pc,Satz2007} that the regularisation~\eqref{Wightman} may be used so long as the switching functions $\chi(\tau)$ are assumed initially to be smooth and of compact support with a sharp-switching limit to be taken carefully later on. While the response function~\eqref{eq:response} diverges at this limit, the rate of response remains finite so that it is this \emph{transition rate}, which we interpret as the rate of particle detection, that will be the quantity of interest. It was shown in Ref.~\cite{Satz2007} (and reproduced for FLRW spacetimes in Appendix~\ref{sec:Sharp}) that the sharp-switching limit of the transition rate agrees with Schlicht's regularisation to leading order and is given by
\begin{align}
	\label{transitionsharp}
	{\cal \dot{F}}_{\tau}(\omega)=\frac{1}{2\pi^{2}}\int_{0}^{\Delta\tau}ds\left(\frac{\cos\omega s}{\sigma^2(\tau,s)}+\frac{1}{s^{2}}\right)+\frac{1}{2\pi^{2}\Delta\tau}-\frac{\omega}{4\pi},
\end{align}
where $\Delta\tau\equiv\tau-\tau_0$ is the detection time and the trajectories\footnote{Explicitly, $(\Delta x)^2=-(\eta(\tau)-\eta(\tau-s))^2+(\mathbf{x}(\tau)-\mathbf{x}(\tau-s))^2$.} $\Delta x\equiv x(\tau)-x(\tau-s)$ are encoded in the geodesic distance $\sigma^2(\tau,s)\equiv a(\tau)a(\tau-s)(\Delta x)^{2}$.

	To track the thermal behaviour of an Unruh-DeWitt detector,	
	we model the temperature of the detector on the detailed balance form of the Kubo–Martin–Schwinger (KMS) conditions, Ref.~\cite{Kubo}. To arrive at a suitable temperature estimator we first define the excitation to de-excitation ratio 
\begin{equation}
\bar{	\cal R} \equiv \frac{\cal F(\omega)}{\cal F(-\omega)}
\end{equation} 
and note that if there exists a constant $T$ that satisfies the detailed-balance form of the KMS condition
\begin{equation}
\bar{\cal R}=e^{-\omega /T},    
\end{equation}
then we identify $T=\bar{T}_{\textrm{EDR}}$ with the temperature of the detector which is given by 
\begin{align}
	\label{eq:TEDR0}
	\bar T_{\textrm{EDR}}=-\frac{\omega}{\ln\mathcal{\bar R}}.
\end{align}

For a static detector coupled to a scalar field in the Hartle-Hawking state, this temperature is independent of the energy gap and equals the locally-measured Hawking temperature in the limit of infinite detection time. For finite detection times, $\bar T_{\textrm{EDR}}$ becomes dependent on the energy gap but this dependence can be sufficiently weak, so that it remains a suitable temperature estimator for the detector, see  Ref.~\cite{Garay2016}.

A complication arises from the fact that the response function diverges in the sharp-switching limit. In this case, instead of the ratio of excitation to de-excitation probabilities as above, we take instead the ratio of the rates 
\begin{align}
	T_{\textrm{EDR}}=-\frac{\omega}{\ln{\mathcal{R}}},\qquad {\mathcal{R}}=\frac{\dot{\mathcal{F}}_{\tau}(\omega)}{\dot{\mathcal{F}}_{\tau}(-\omega)}.
	\label{TEDR}
\end{align}
As shown in Refs.~\cite{Conroy13,HodgkinsonLoukoOttewill}, this definition  gives the expected $T_{\textrm{EDR}}=T_{\textrm{loc}}$ in the limit of infinite detection time for a static detector coupled to a field in the Hartle-Hawking state, where $T_{\textrm{loc}}$ is the red-shifted Hawking temperature of the black hole.  Similarly, a geodesic  detector coupled to a field in the conformal vacuum of the de Sitter universe also gives the expected $T_{EDR}=T_{dS}^{\textrm{loc}}$, while a comoving detector simply gives $T_{EDR}=T_{dS}$. We show this explicitly in Section \ref{sec:dS}. In each case, Eq.~(\ref{TEDR}) remains a suitable temperature estimator for finite but sufficiently long times in the sense that the dependence on $\omega$ is weak and $T_{\textrm{EDR}}$ asymptotes to the locally measured field temperature as the detection time is increased. The temperature estimator Eq.~(\ref{TEDR}) has been employed in Ref.~\cite{HodgkinsonLoukoOttewill} for a detector on a circular geodesic in Schwarzschild in the limit of infinite detection time, and in the context of the near-horizon regime of an extremal black hole in Ref.~\cite{Conroy13}. 

\section{Surface gravity on a dynamical horizon}
\label{sec:SurfaceGravity}
\subsection{Surface gravity on a Killing horizon}
To better understand the concept of surface gravity on a dynamical horizon it is prudent to first review the situation in a static or stationary spacetime which allows for the existence of a Killing vector $\chi^\mu$, satisfying the Killing equation $\nabla_{(\mu}\chi_{\nu)}=0$. Indeed, a timelike Killing vector may only persist in a stationary spacetime. In the region where $\chi^\mu$ is timelike, the quantity $\chi^{\sigma}\chi_{\sigma}$ is negative, while a Killing horizon is formed on the surface where the Killing vector becomes null and $\chi^{\sigma}\chi_{\sigma}=0$. In the context of a static black hole, we interpret this null surface as the event horizon and the timelike region as the exterior. 

More generally, a Killing horizon is a null hypersurface defined by a Killing vector with vanishing norm meaning that a Killing vector is, by definition, normal to the horizon. Indeed, it is a somewhat counter-intuitive property of null surfaces that any null vector that is normal to a null surface will also be tangent to it. Thus, as the norm $\chi^{\sigma}\chi_{\sigma}$ vanishes everywhere on the horizon, the gradient $\nabla_{\mu}(\chi^{\sigma}\chi_{\sigma})$ will be directed along $\chi_\mu$. That is to say that  $\chi_\mu$ and  $\nabla_{\mu}(\chi^{\sigma}\chi_{\sigma})$ are proportional to each other, Ref.~\cite{Poisson:2009pwt}. Thus, with the benefit of hindsight, we may write
\begin{equation}
	\label{Killing0}
	\nabla_{\mu}(\chi^{\sigma}\chi_{\sigma})=-2\kappa\chi_{\mu}\quad\implies\quad\chi^{\sigma}\nabla_{\mu}\chi_{\sigma}=-\kappa\chi_{\mu},
\end{equation}
for some scalar function $\kappa$. A straightforward application of the Killing equation reveals 
\begin{equation}
	\chi^{\sigma}\nabla_{\sigma}\chi^{\mu}=\kappa\chi^{\mu},
\end{equation}
and we interpret $\kappa$ to be the surface gravity on the Killing horizon.  Equivalently, we may write
\begin{equation}
\kappa^{2}=V^{-2}(\chi^{\sigma}\nabla_{\sigma}\chi^{\mu})(\chi^{\lambda}\nabla_{\lambda}\chi_{\mu}),
\end{equation}
where $V=\sqrt{|\chi^{\mu}\chi_{\mu}|}$ is the red-shift factor, Refs. \cite{Carroll:2004st,Wald:GR}.

It is here that the relationship with acceleration becomes apparent. If we write the four-acceleration $A^{\mu}\equiv u^{\sigma}\nabla_{\sigma}u^{\mu}$ in terms of the red-shift factor $V$ and the Killing vector $\chi^\mu$ using $u^\mu=\chi^\mu/V$, we may rearrange the above equation and evaluate on the horizon to yield~
\begin{equation}
	\label{VAstatic}
	\kappa\; \big|_{r=r_H}=V A\;\big|_{r=r_{H}},
\end{equation}
where $A=\sqrt{|A^\mu A_\mu|}$ is the magnitude of the four-acceleration.  We interpret $A$ as the locally-applied force required to hold a particle in position at some radius $r$ which, predictably, diverges on the horizon $r_{H}$. The redshift factor serves to shift the application of this force to infinity so that the interpretation of $\kappa$ is the gravitational force (acceleration) that must be applied in order to hold a particle in place near the horizon (i.e. the surface gravity), where this force is not locally-applied but applied at infinity. This ensures that the Killing surface gravity $\kappa$ is regular when evaluated on the horizon while also demonstrating the non-local nature intrinsic to this definition, Refs.~\cite{Poisson:2009pwt,Nielsen:2007ac,Faraoni:2015ula}. 

\subsection{Hayward-Kodama surface gravity}
Of course, the above interpretation of surface gravity rests on the existence of a Killing vector which are only admitted in stationary (and static) spacetimes. Consequently, as dynamical spacetimes don't allow timelike Killing vectors, physical quantities such as the four-acceleration and the surface gravity on an evolving horizon become ambiguous, Ref.~\cite{Abreu:2010ru}. The motivation then is to define a vector in a dynamical spacetime which describes a symmetry of spacetime in a geometric- and coordinate-independent way in order to lend meaning to these physical quantities just as the Killing vector does for stationary spacetimes. The insight of Hideo Kodama in Ref.~\cite{Kodama} was to note that there exists a divergence-free vector field $k^a$ for any spherically-symmetric, time-dependent metric which defines a class of preferred observers $u^a\equiv k^a/\sqrt{|k^ck_c|}$ in the region where it is timelike, \cite{Faraoni:2015ula,Kodama}.

We call such a vector a \emph{Kodama vector} (KV) and define it via\footnote{Our convention throughout is that Greek indices $\mu,\nu,\dots$ run from $0,1,2,3$, while Latin indices towards the start of the alphabet $a,b,\dots$ represent the first two indices $0,1$, and indices towards the middle $i,j,\dots$ indicate spatial component $1,2,3$.} 
\begin{equation}
	\label{Kodamadef}
	k^{a}\equiv\epsilon^{ab}\nabla_{b}\tilde{r},\quad\text{with }\quad k^{\theta}=k^{\varphi}=0,
\end{equation}
where $\tilde r$ is the radial coordinate of the spherically, symmetric dynamical spacetime, $\epsilon^{ab}$ is the $(1+1)$ Levi-Civita tensor and $a,b=\{t,r\}$. In the context of an FLRW metric, which is our focus here, the radial coordinate $\tilde r\equiv a(t)r$ is the areal radius. As the KV is divergence-free (see Ref.~\cite{Abreu:2010ru} for further details), the expansion $\Theta\equiv \nabla_\mu k^\mu$ necessarily vanishes. Thus, from the point of view of a Kodama observer, the dynamical, spherically-symmetric background will appear not to expand.  Intuitively, one can then surmise that the areal radius of the spacetime remains constant so that, for a Kodama observer, the radial coordinate is defined as 
\begin{equation}
	\label{radialKod0}
	r=\frac{K}{a},\quad \mbox{for some constant }K.
\end{equation}
The divergence-free nature of the KV means that one can define a current $J^a\equiv G^{ab}k_b$, where $G^{ab}$ is the Einstein tensor, which is covariantly conserved. The unexpected existence of this conserved current is sometimes referred to as the \emph{Kodama miracle}, Refs.~\cite{Abreu:2010ru,Kodama,Faraoni:2015ula}.

Following the same prescription as in the static case, we consider a Kodama vector $k_a$ along with the vector field $\nabla_a(k^c k_c)$ which are both normal to some trapping surface or apparent horizon\footnote{See Appendix~\ref{sec:Horizons} for a note on the terminology used here.} which we think of as analogous to the the null hypersurface in the stationary example. 
 As these two quantities are proportional to each other we find
\begin{equation}
	\label{KodamaSG0}
	k^c \nabla_a k_c=\kappa_{HK} k_a,
\end{equation}
which is comparable to the second identity in Eq.~\eqref{Killing0}, where we have chosen the sign to reflect that we intend to evaluate the surface gravity on the cosmological apparent horizon defined in Eq.~\eqref{rAH} and not some marginally outer-trapped surface found in dynamical black hole spacetimes. In general, we can not make use of the Killing equation as we did at this point in the stationary case and should instead deploy the altered form 
\begin{equation}
k^a\left(\nabla_a k_b+\nabla_b k_a\right)=8\pi G \tilde r \psi_b,
\end{equation}
which tracks the deviation of the Kodama vector from a Killing vector. Here we have introduced the energy flux vector $\psi_a$, the details of which can be found in Refs.~\cite{Helou:2014ela,Hayward:1997jp}. If we now stipulate that the Kodama vector conforms to the Killing equation by choosing $\psi_a=0$ as in Ref.~\cite{Hayward:1997jp}, we ensure that the surface gravity is uniquely defined and recovers the Reissner-Nordstrom surface gravity in the static limit.
A consequence of this is that the Kodama trajectory is no longer geodesic so that a detector travelling along such a trajectory will require some acceleration. With this in mind, Eq.~\eqref{KodamaSG0} becomes
\begin{equation}
	\frac{1}{2}g^{ab}k^{c}(\nabla_{c}k_{a}-\nabla_{a}k_{c})=-\kappa_{HK}k^{b}.
\end{equation}
By decomposing a generic, spherically symmetric spacetime metric into the form
\begin{equation}
	\label{gammametric}
	ds^2=\gamma_{ab} dx^a dx^b+\tilde r^2 d\Omega^2,
\end{equation}
where $\gamma_{ab}$ is a $2$-dimensional metric and all the spherical coordinates are contained in the $2$-sphere $d\Omega^2$, we can write $\kappa_{HK}$ in the geometric form
\begin{equation}
	\label{surfaceHK}
\kappa_{HK}=-\frac{1}{2}\Box_{(\gamma)}\tilde{r}=-\frac{1}{2}\frac{1}{\sqrt{-\gamma}}\partial_{a}\left(\sqrt{-\gamma}\gamma^{ab}\partial_{b}\tilde{r}\right),
\end{equation}
which defines the \emph{Hayward-Kodama surface gravity}. 

As before, we can express Eq.~\eqref{surfaceHK} in terms of the magnitude of the four-acceleration $A^a\equiv u^c \nabla_c u^a=V_k^{-2} k^c\nabla_c k^a$ (where $V_k=\sqrt{|k^c k_c|}$ is the red-shift factor) like so
\begin{equation}
	\kappa_{HK}=V_k  A.
\end{equation}
When evaluated on an appropriate horizon, this is of precisely the same form as Eq.\eqref{VAstatic} so that our prior interpretation of $\kappa_{HK}$ as surface gravity is retained.

The horizon we are interested in is the cosmological apparent horizon which occurs when the ingoing expansion vanishes, as discussed in Appendix \ref{sec:Horizons}. We take as our background metric the line element for FLRW in comoving coordinates
\begin{equation}
	\label{FLRW}
	ds^2 =-dt^2 +a^2 (t)\left(dr^2+r^2d\Omega^2\right),
\end{equation} 
where $d\Omega^2$ is the line element for the $2$-sphere and the scale factor $a(t)$ parametrises the evolution of the universe. In this case, we find the surface gravity to be 
\begin{equation}
	\label{kappaH0}
	\kappa_{HK}=\tilde{r}\left(H^{2}+\frac{1}{2}\dot{H}\right),
\end{equation}
where $H\equiv\frac{\dot a}{a}$ is the Hubble parameter. Recall that $\tilde r \equiv a(t) r$ signifies the areal radius so that $\tilde r_{AH}$ is the areal radius evaluated on the horizon $r=r_{AH}$. Expressing Eq.~\eqref{kappaH0} in terms of this horizon using $\tilde r_{AH} = 1/H$ and evaluating on $\tilde r=\tilde r_{AH}$ then yields
\begin{equation}
		\label{kappaHay}
	\kappa_{HK}\big|_{\tilde r=\tilde r_{AH}}=	\frac{1}{\tilde{r}_{AH}}\left[1-\frac{\dot{\tilde{r}}_{AH}}{2H\tilde{r}_{AH}}\right],
\end{equation}
where\footnote{Note that this is the general form for $\kappa_{HK}$ and $\dot{\tilde{r}}_{AH}$ in that Eqs.~\eqref{kappaHay}--\eqref{THK} are valid also for FRW spacetimes with non-flat geometry. In the spatially-flat case, which is our focus here, we have the added simplification that $\tilde r_{AH}=1/H$.}
\begin{equation}
	\label{kappaHay2}
	\tilde{r}_{AH}=H^{-1}\quad\implies\quad\dot{\tilde{r}}_{AH}=-\tilde{r}_{AH}^{3}H{\dot H}.
\end{equation}
This is indeed regular on the cosmological apparent horizon in the case considered while, more generally, the Hayward-Kodama surface gravity defined in Eq.~\eqref{surfaceHK} has been shown in Ref.~\cite{Nielsen:2007ac} to give the correct Killing horizon behaviour in the static limit and, in Ref.~\cite{Hayward:1997jp}, to conform to a consistent `Unified first law of thermodynamics'. With this definition of surface gravity comes the associated temperature
\begin{equation}
	\label{THK}
	T_{HK}=	\frac{1}{2\pi\tilde{r}_{AH}}\left[1-\frac{\dot{\tilde{r}}_{AH}}{2H\tilde{r}_{AH}}\right].
\end{equation}
\subsection{Effective surface gravity.}
An alternative formulation was put forward by Ashtekar et al.~in Ref.~\cite{Ashtekar:2004cn} which we refer to as \emph{effective surface gravity}. To arrive at this definition we begin by defining the scalar quantity 
\begin{equation}
	\label{chidef}
	\chi\equiv\gamma^{ab}\partial_{a}\tilde{r}\partial_{b}\tilde{r},
\end{equation}
where $\gamma$ is the two-dimensional metric defined in Eq.~\eqref{gammametric} and the
indices $a.b$ take the values of $0$ and $1$. For the metric in comoving form \eqref{FLRW}, we have $\chi=1-H^2 \tilde r^2$
which can be written as
\begin{equation}
	\chi=1-(\tilde{r}/\tilde{r}_{AH})^{2}.
\end{equation}
The \emph{effective surface gravity} and associated temperature  on the dynamical horizon  are defined by
\begin{equation}
	\label{effectivesurface}
	\kappa_{eff}  \equiv-\frac{1}{2}\frac{\partial\chi}{\partial\tilde{r}}\biggr|_{\tilde{r}=\tilde{r}_{AH}}=\frac{1}{\tilde{r}_{AH}}\quad \mbox{and} \quad 	T_{eff}\equiv \frac{1}{2\pi\tilde{r}_{AH}}.
\end{equation}
This definition displays the familiar form of a spherically-symmetric black hole and preserves the correct Killing horizon behaviour of the Schwarzschild metric in the static limit, whilst also providing a local interpretation of surface gravity. 
\subsection{Fodor surface gravity}
Recall that the minimally anti-trapped surface occurs when the ingoing expansion \eqref{expin} vanishes and the outgoing expansion remains positive. This is a `marginal' surface in that the expansion vanishes and `minimal' in that it is the anti-trapped surface of smallest physical size. Within the context of an FLRW spacetime, we interpret this surface as the cosmological apparent horizon. However, in a generic dynamical spacetime, we may, in principle, also consider the surface formed when the outgoing expansion vanishes and the ingoing expansion is negative. We call such a surface \emph{marginally outer-trapped}, Ref.~\cite{Nielsen:2007ac}. 

The standard Fodor approach to surface gravity, first put forward in Ref.~\cite{Fodor:1996rf}, relies on the presence of such a marginally outer-trapped surface. It is defined via the identity $\kappa l^a=l^b \nabla_b l^a$  where, as in Eq.~\eqref{inoutrays}, $l^a$ are outgoing  null tangent vectors. While marginally outer-trapped surfaces may be a significant feature of dynamical black holes, our touchstone, in the cosmological context, is the cosmological apparent horizon which occurs when the \emph{ingoing} expansion vanishes. We therefore propose an altered form of Fodor surface gravity by simply replacing the outgoing tangent vector $l^a$ with the ingoing vector $\bar n^a$ like so
\begin{equation}
	\label{Fodor}
	\kappa_F \bar n^a= \bar n^b \nabla_b \bar n^a.
\end{equation}
This alteration allows us to evaluate the Fodor surface gravity on the cosmological apparent horizon and to remain consistent with our previous formulations of surface gravity. 

To find an explicit expression for $\kappa_F$ in an FLRW spacetime, we first rewrite Eq.~\eqref{FLRW} in Painlev\'e-Gullstrand (PG) coordinates and verify that the null tangent vectors 
\begin{equation}
	\label{PGnullrays2}
	\bar n^\mu=\left(1,-1+H\tilde r,0,0\right),\quad 	\bar l^\mu=\left(1,1+H\tilde r,0,0\right),
\end{equation}
have a cross normalisation of $\bar n^a \bar l_a=-2$. Further details can be found in Appendix \ref{sec:Fodorapp}. Keeping with PG coordinates, we unbox Eq.~\eqref{Fodor} to obtain
\begin{equation}
	\kappa_{F}=\frac{1}{2}H\left(2-(\tilde r H)+(\tilde r H)^2\right),
\end{equation}
which can be written in terms of $\tilde r_{AH}=1/H$ like so
\begin{equation}
	\kappa_{F}=\frac{1}{2\tilde r_{AH}}\left(2-(\tilde r /\tilde r_{AH})+(\tilde r/\tilde r_{AH})^2\right).
\end{equation}
Finally, we evaluate on the cosmological apparent horizon to find
\begin{equation}
	\kappa_{F}\big|_{\tilde{r}=\tilde{r}_{AH}}=\frac{1}{\tilde{r}_{AH}}.
\end{equation}
We see here that our cosmological reformulation of Fodor surface gravity produces the same result as  effective surface gravity given in Eq.~\eqref{effectivesurface} when evaluated on the cosmological apparent horizon.
\subsection{Nielsen-Visser surface gravity}
As with $\kappa_{eff}$, our starting point for Nielsen-Visser surface gravity is to consider the scalar quantity given in Eq.~\eqref{chidef}, rearranged in terms of the Misner-Sharp-Hernandez (MSH) mass $M=M(t,\tilde{r})$, like so
\begin{equation}
	\chi\equiv\gamma^{ab}\partial_a \tilde r \partial_b \tilde r \equiv 1-\frac{2M}{\tilde r},
\end{equation}
where the Nielsen-Visser surface gravity is given by
\begin{equation}
	\label{NVsurfacegrav}
	\kappa_{NV}\equiv - \frac{1-2M^\prime}{2\tilde{r}},
\end{equation}
and the prime $^\prime$ indicates a derivative w.r.t. the areal radius $\tilde r$. The overall minus sign comes from the direction of travel in the cosmological context, Ref.~\cite{Nielsen:2005af}. This would be omitted in the case of a dynamical black hole. To compute the explicit value on the cosmological apparent horizon, we first recast the PG form of the FLRW metric \eqref{FLRWPG} in terms of the MSH mass
\begin{equation}
	ds^2=-\left(1-\frac{2M}{\tilde r}\right)dt^2-2\sqrt{\frac{2M}{\tilde r}}dt d\tilde r+d\tilde{r}^2+\tilde{r}^2 d\Omega^2,
\end{equation}
and compute Eq.~\eqref{NVsurfacegrav} to obtain
\begin{equation}
\kappa_{NV}=-\frac{1-3(\tilde{r}/\tilde{r}_{AH})^{2}}{2\tilde{r}}.
\end{equation}
When evaluated on $\tilde r=r_{AH}$, we again find
\begin{equation}
	\kappa_{NV}|_{\tilde{r}=r_{AH}}=\frac{1}{\tilde{r}_{AH}}.
\end{equation}

In summary, the approaches of Nielsen-Visser, Fodor and Ashtekar, while distinct a priori, align when evaluated on the cosmological apparent horizon of a geometrically-flat FLRW universe. As such, for the remainder of the article we group these approaches together under the umbrella of `effective surface gravity' to compare with the approach of Hayward-Kodama in Eq.~\eqref{surfaceHK}.
\section{Trajectory}
\label{sec:Trajectory}
\subsection{Geodesic trajectories}
Our background metric is that of the FLRW metric in comoving coordinates given by Eq.~\eqref{FLRW}. For geodesic motion, we compute the four-velocity in the usual manner to find
\begin{equation}
	u^{\mu}=\left(\sqrt{1+\frac{c_{1}^{2}}{a^{2}}},\frac{c_{1}}{a^{2}},0,0\right),
\end{equation}
where $c_1$ is a constant. This encodes the geodesic trajectories $x^\mu(\tau)=\int u^\mu(\tau)d\tau$ that the detector will follow, where $\tau$ is the proper time experienced by the detector. Along with generic geodesic motion, we are interested in the special case of the comoving observer which is characterised by the choice of $c_1=0$. Explicitly, the four-velocity in each case is given by
\begin{equation}
	\label{geodesicU}
	u_{\text{comoving}}^{\mu}=\left(1,0,0,0\right),\quad u^{\mu}=\left(\sqrt{1+\frac{c_1^2}{a^{2}}},\frac{c_1}{a^{2}},0,0\right).
\end{equation}
In conformal coordinates, the metric \eqref{FLRW} becomes
	\begin{equation}
	\label{FLRWconf}
	ds^2=a^2(\eta)\left(-d\eta^2+dr^2+r^2d\Omega^2\right),
\end{equation}
where we have introduced the conformal time coordinate $\eta$ via $d\eta=a^{-1}dt$, which yields the four-velocities
\begin{equation}
	\label{trajgeoconf}
	v_{\text{comoving}}^{\mu}=\left(\frac{1}{a},0,0,0\right),\quad v^{\mu}=\frac{1}{a}\left(\sqrt{1+\frac{c_{1}^{2}}{a^{2}}},\frac{c_{1}}{a},0,0\right).
\end{equation}
Thus, the non-vanishing trajectories for a comoving observer are given by
\begin{equation}
	\label{trajcom}
	\eta (\tau)=\int\frac{d\tau}{a},\quad r=r_0,
\end{equation}
while for a generic geodesic observer, we have
\begin{equation}
	\eta(\tau)=\int \frac{1}{a}\sqrt{1+\frac{c_1^2}{a^2}}d\tau,\quad r(\tau)=\int \frac{c_1}{a^2}d\tau.
	\label{geotrajc}
\end{equation}
\subsection{Kodama trajectories}
We will also consider non-geodesic motion in the form of a detector accelerating along a Kodama trajectory. To write down the explicit form of this trajectory in FLRW, we begin with a generic, spherically-symmetric, four-dimensional spacetime metric $g$ decomposed as in Eq.~\eqref{gammametric}. From the definition given in Eq.~\eqref{Kodamadef}, the Kodama vector for the above metric is given by
\begin{equation}
	k^{a}=\sqrt{-h}\left(a\gamma^{at}\gamma^{rr}-\dot{a}r\gamma^{ar}\gamma^{tt}\right).
\end{equation}

Assuming a geometrically-flat FLRW metric in comoving coordinates as in Eq.~\eqref{FLRW}, the non-vanishing components of the Kodama vector are given by 
\begin{equation}
	k^{0}=-1,\quad k^{1}=Hr,\quad k^{c}k_{c}=-1+\dot{a}^{2}r^{2},
\end{equation}
where $H\equiv H(t)$ and the final term is closely related to the red-shift factor $V_k\equiv\sqrt{|k^ck_c|}$. Indeed, we can rewrite the latter identity in terms of the apparent horizon $r_{AH}$ like so 
\begin{equation}
	k^{c}k_{c}=-1+( r/r_{AH})^{2}.
		\label{kckc_hor}
\end{equation}
 In this form, it is clear to see that the Kodama vector does indeed mimic the Killing vector in that it becomes null on the surface of the apparent horizon $r=r_{AH}$ and is timelike in the region $r<r_{AH}$. In the region where it is timelike, the Kodama vector evokes a class of preferred observers with four-velocity $u^a\equiv k^a/V_k$, given by
\begin{equation}
	\label{fourvel0}
	u^{\mu}=\frac{1}{\sqrt{1-\dot{a}^{2} r^{2}}}\left(-1,Hr,0,0\right).
\end{equation}

If we now assume the conformal form of the metric~\eqref{FLRWconf}, we can read off the explicit components
\begin{equation}
	k^{0}=-\frac{1}{a},\qquad k^{1}=\frac{{\cal H}r}{a},\qquad k^{c}k_{c}=-1+{\cal H}^{2}r^{2},
\end{equation}
where  ${\cal H}\equiv a^\prime/a$ is the Hubble parameter in conformal coordinates and $^\prime$ indicates a derivative w.r.t. conformal time $\eta$. The apparent horizon in this case is given by $r_{AH}={\cal H}^{-1}$ and conforms to the same relation given in Eq.~\eqref{kckc_hor}, while the four-velocity is found to be
\begin{equation}
	\label{fourvel}
	v^{\mu}=\frac{1}{a\sqrt{1-{\cal H}^{2}r^{2}}}\left(-1,{\cal H}r,0,0\right).
\end{equation}
With $v^0=d\eta/d\tau$ and $v^1=dr/d\tau$, we integrate to find the trajectories 
\begin{equation}
	\label{traj0}
	\eta(\tau)=\int\frac{\sqrt{V(\tau)}}{a}d\tau,\qquad r(\tau)=\frac{K}{a},
\end{equation}
where we have defined $V(\tau)\equiv1+K^2 H^2(\tau)$ and $H(\tau)\equiv{\mathring a}/a$ with~$\mathring{~}$~signifying a derivative w.r.t. proper time. We observe here that the radial coordinate agrees with our earlier intuition in Eq.~\eqref{radialKod0} that the areal radius for a Kodama observer is constant.

\section{A detector in the de Sitter universe}
\label{sec:dS}
Consider a transition rate of the form 
\begin{equation}
{\cal \dot{F}}_{\tau}(\omega)=\frac{1}{2\pi^{2}}\int_{0}^{\Delta\tau}ds\left(\frac{\cos\omega s}{\sigma^{2}(\tau,s)}+\frac{1}{s^{2}}\right)+\frac{1}{2\pi^{2}\Delta\tau}-\frac{\omega}{4\pi},
\label{TranRateSharp}
\end{equation}
where $\sigma^{2}(\tau,s)\equiv a(\tau)a(\tau-s)(\Delta x)^{2}$ is
the geodesic distance and $(\Delta x)^{2}\equiv\eta_{\mu\nu}\Delta x^{\mu}\Delta x^{\nu}$.
Equivalently, we may write
\begin{equation}
	\label{dSRate0}
{\cal \dot{F}}_{\tau}(\omega)=\frac{1}{2\pi^{2}}\int_{0}^{\infty}ds\left(\frac{\cos\omega s}{\sigma^{2}(\tau,s)}+\frac{1}{s^{2}}\right)+J_{\tau}-\frac{\omega}{4\pi},
\end{equation}
where we define the `fluctuating tail' 
\begin{equation}
J_{\tau}\equiv-\frac{1}{2\pi^{2}}\int_{\Delta\tau}^{\infty}\frac{\cos\omega s}{\sigma^{2}(\tau,s)}ds,
\end{equation}
which vanishes at the limit $\Delta\tau\to\infty$. Next, by adding
and subtracting a $\cos(\omega s)/s^{2}$ term in the integrand, we
can re-write Eq.~\eqref{dSRate0} like so
\begin{align}
{\cal \dot{F}}_{\tau}(\omega)&=\frac{1}{2\pi^{2}}\int_{0}^{\infty}ds\cos\omega s\left(\frac{1}{\sigma^{2}(\tau,s)}+\frac{1}{s^{2}}\right)
\nn\\&+J_{\tau}+\frac{|\omega|-\omega}{4\pi},
\end{align}
The final term is of precisely the same form as an inertial detector
coupled to a field in the Minkowski vacuum \cite{HodgkinsonLoukoOttewill}, so that we may split the transition rate into its inertial portion and the non-inertial correction like so
\begin{equation}
	\label{Ratedecomp}
{\cal \dot{F}}_{\tau}(\omega)={\cal \dot{F}}_{\tau}^{\text{corr}}(\omega)+{\cal \dot{F}}_{\tau}^{\text{inertial}}(\omega),
\end{equation}
where
\begin{equation}
	\label{Ratecorr}
{\cal \dot{F}}_{\tau}^{\text{corr}}(\omega)\equiv\frac{1}{2\pi^{2}}\int_{0}^{\infty}ds\;\cos\omega s\left(\frac{1}{\sigma^{2}(\tau,s)}+\frac{1}{s^{2}}\right)+J_{\tau}.
\end{equation}
De Sitter space in FLRW coordinates is characterised by the scale factor $a(t)=e^{Ht}$, where
$H$ is now the Hubble constant. This ensures that the geodesic distance
is independent of proper time, i.e. $\sigma^{2}(\tau,s)=\sigma^{2}(s)$,
a simplification which is present also in the case of stationary (and static) black
hole spacetimes. 
\subsection{Comoving detector}
For a comoving detector, cosmic time and proper time are interchangeable, i.e. $t=\tau$,
so that the scale factor in terms of proper time is simply $a(\tau)=e^{H\tau}$.
It is then straightorward to compute the conformal time trajectory
\eqref{trajcom} and the geodesic distance which we find to be 
\begin{equation}
\eta(\tau)=-H^{-1}e^{-H\tau},\quad\sigma^{2}(s)=-4H^{-2}\sinh^{2}\left(\frac{Hs}{2}\right).
\end{equation}
The simplicity of these components allows us to compute the transition
rate analytically. To this end, we rewrite Eq.~\eqref{Ratecorr} like so 
\begin{equation}
	\label{FcorrdScom}
{\cal \dot{F}}_{\tau}^{\text{corr}}(\omega)\equiv\frac{H}{8\pi^{2}}\int_{-\infty}^{\infty}ds\;e^{-2i\omega s/H}\left(-\frac{1}{\sinh^{2}s}+\frac{1}{s^{2}}\right)+J_{\tau},
\end{equation}
where we have used the fact that the integrand is an even function
to extend the interval and have introduced the change of variables
$s\to2s/H$. Evaluating this using contour integration and the theory
of residues gives 
\begin{equation}
{\cal \dot{F}}_{\tau}(\omega)\equiv\frac{\omega}{2\pi}\left[\frac{1}{e^{2\pi\omega/H}-1}\right]+J_{\tau},
\end{equation}
where the fluctuating tail $J_{\tau}$ is evaluated in terms of incomplete
Beta function like so 
\begin{align}
	\label{dSTailExplicit}
\nn	J_{\tau} & =\frac{i\omega}{8\pi^{2}}\biggl[B(e^{-H\tau};1+\frac{i\omega}{H},0)-B(e^{-H\tau};1-\frac{i\omega}{H},0)\\
	& +B(e^{-H\tau};\frac{i\omega}{H},0)-B(e^{-H\tau};-\frac{i\omega}{H},0)\biggr]
	\nn\\&-\frac{H}{4\pi^{2}}\coth\left(\frac{H\tau}{2}\right)\cos(\tau\omega),
\end{align}
which is valid\footnote{The expression is in fact valid on the branch of solutions where proper time $\tau$ and the Hubble constant $H$ have the same sign. In the de Sitter universe $H>0$, while proper time is positive so that Eq.~\eqref{dSTailExplicit} is the representative branch of solutions to be considered. An alternate branch of solutions, which may prove useful for an analysis in anti-de Sitter space, can be found where $H$ and $\tau$ come with opposing signs.} for real $\tau>0$ and $H>0$. The tail of course vanishes at the limit of large detection time, leaving
\begin{equation}
	\label{RatedScominf}
\dot{{\cal F}}_{\infty}(\omega)=\frac{\omega}{2\pi}\left(\frac{1}{e^{\omega/T_{dS}}-1}\right),\quad T_{dS}=H/2\pi,
\end{equation}
where $T_{dS}$ is the \emph{de Sitter temperature}, i.e. the temperature
an inertial observer in de Sitter space will read on their thermometer. We see here that
the inertial piece drops out for both positive and negative energy
gaps when integration on the contour is taken carefully, leaving only
the non-inertial, curved space correction. This registers an exactly
Planckian distribution for a blackbody in thermal equilibrium in the
limit of large detection time, mirroring the behaviour of a static
detector coupled to a field in the Hartle-Hawking state in a black
hole spacetime, Ref.~\cite{HartleHawking1976}. In the case of a comoving observer, the KMS temperature
is precisely that of the de Sitter temperature while, in the black hole case, the KMS temperature is
the locally-measured Hawking temperature, Ref.~\cite{Conroy13}.

	\begin{figure}[htb]
	\centering
		\includegraphics[width=\linewidth]{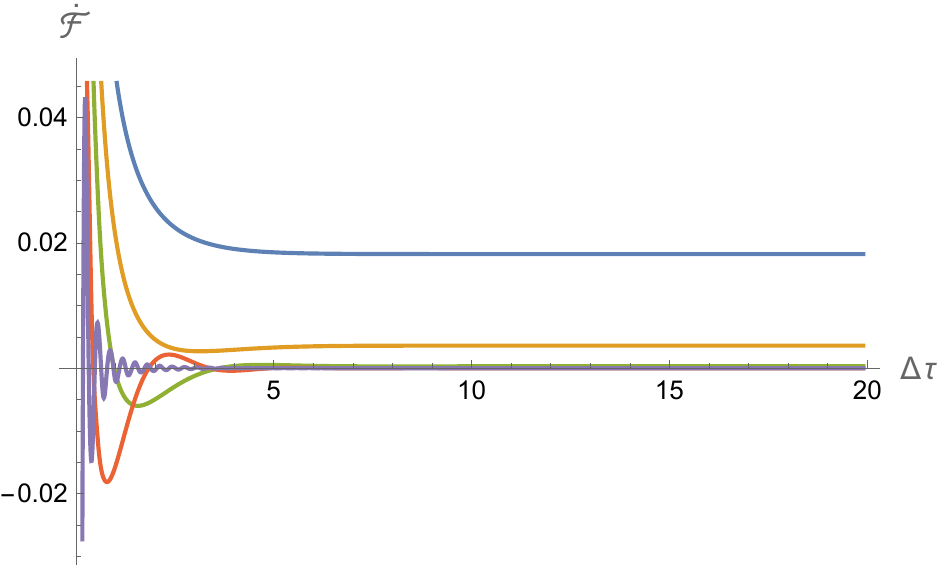}
	\caption{Figure shows the transition rate of a comoving detector in de Sitter space as a function of detection time. A selection of energy gaps are depicted, i.e. $\omega=1/10$ (blue), $\omega=1/2$ (yellow), $\omega=1$ (green), $\omega=2$ (red), and $\omega=20$ (purple). In each case, we have set the Hubble constant to $H=1$.}
	\label{fig:RatedScom_tau}
\end{figure}
Having established that a comoving detector reaches thermal equilibrium in the de Sitter universe  at the limit of large detection time, we turn our attention now to finite values of detection time. In Fig.~\ref{fig:RatedScom_tau}, we track how the transition rate develops over time for a selection of energy gaps. The overall trend here is that  transient oscillations associated with turning the detector on sharply are present when the detection time is small and damped with increased detection time. The frequency of these oscillations increases with $\omega$ so that when the energy gap is small (see blue curve with $\omega=1/10$) the oscillations present themselves as a `dip' in transition rate. As the energy gap increases the oscillatory behaviour becomes more apparent, best exemplified by the purple curve with $\omega=20$. 

As these oscillations are clear signals of transient behaviour, we have in mind a detector which is dominated by transience for short detection times and asymptotes to an approximately constant value as the detection time increases. To understand this further, we move now to the temperature of the detector, which we interpret via the temperature estimator $T_{EDR}$ defined in Eq.~\eqref{TEDR}.

	\begin{figure}[htb]
	\centering
		\includegraphics[width=\linewidth]{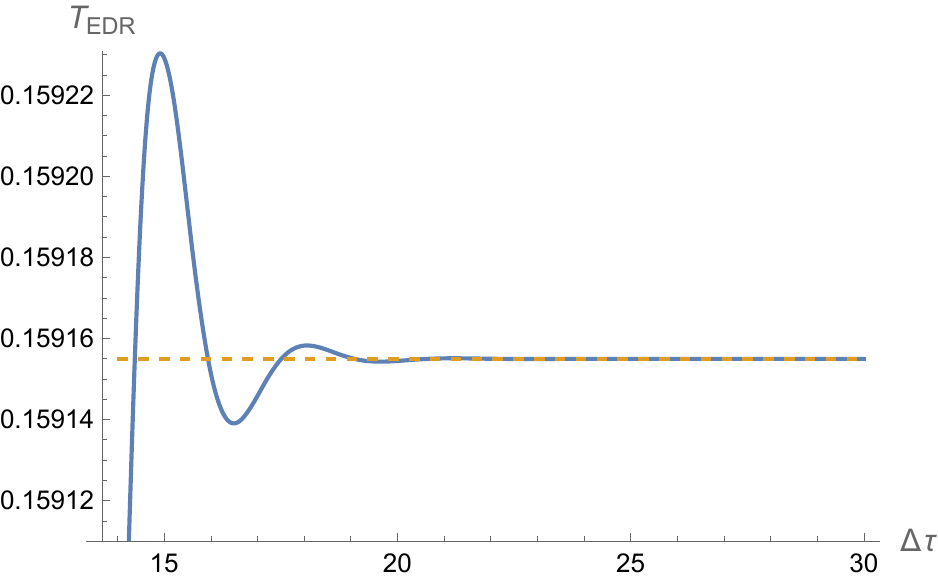}
	\caption{Figure shows the temperature estimator $T_{EDR}$ defined in Eq.~\eqref{TEDR} as a function of detection time for a comoving detector in the de Sitter universe. Early transient oscillations give way to a constant value which coincides with the de Sitter temperature $T_{dS}$, depicted here by the yellow, dashed line. Here we have chosen $\omega=2$ and $H=1$.}
	\label{fig:TEDRcomdS}
\end{figure}
	\begin{figure}[htb]
	\centering
	\includegraphics[width=\linewidth]{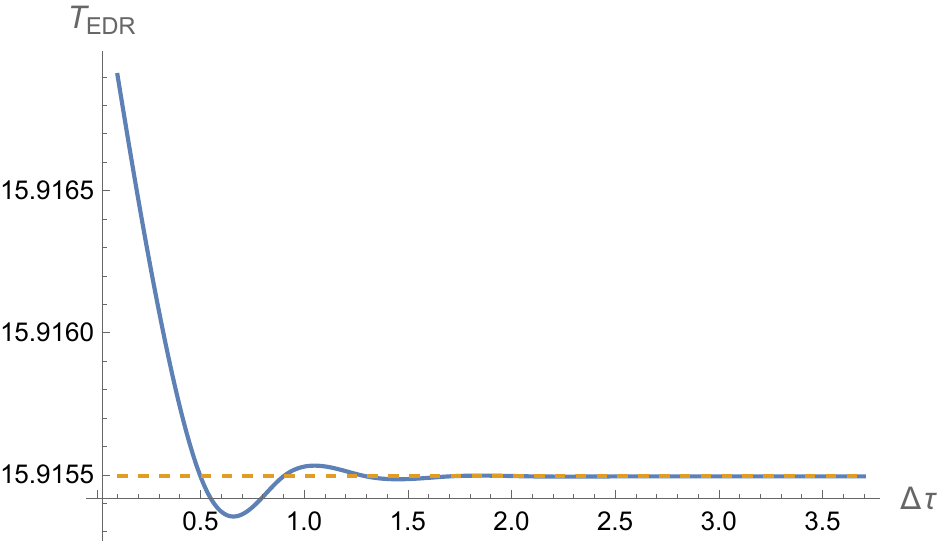}
	\caption{Figure shows the temperature estimator $T_{EDR}$ defined in Eq.~\eqref{TEDR} as a function of detection time for a comoving detector in the de Sitter universe. In comparison with Fig.~\ref{fig:TEDRcomdS}, the de Sitter temperature $T_{dS}\equiv H/2\pi$ (yellow line) has been increased via the choice of $H=100$. As a result, transient oscillations are restricted to earlier time frames and thermalisation is reached sooner. As before, we have chosen $\omega=2$.}
	\label{fig:TEDRcomdS_2}
\end{figure}
We consider this in Fig.~\ref{fig:TEDRcomdS} by plotting $T_{EDR}$ as a function of detection time. Here we observe that (relatively) large, transient oscillations for short detection times dampen over time, before giving way to an approximately constant value after a suitably long detection time. We interpret this constant as the locally-measured KMS temperature or, simply put, the temperature of the detector. Moreover, we see that the temperature of the detector asymptotes to the de Sitter temperature $T_{dS}$ for large detection times, a fact we established analytically in Eq.~\eqref{RatedScominf}. Thus, we can now pin-point more accurately when the detector becomes (approximately) thermal. In the case of Fig.~\ref{fig:TEDRcomdS} we have chosen $\omega=2$ and $H=1$, and find that the detector thermalises to $T_{dS}$ when $\Delta\tau \approx 25$. 

	\begin{figure}[htb]
	\centering
	\includegraphics[width=\linewidth]{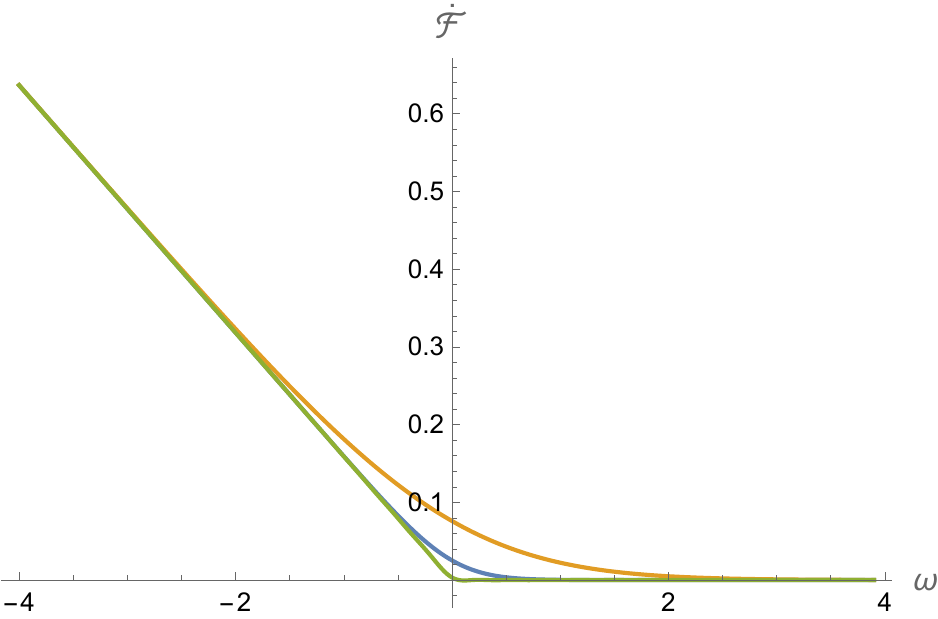}
	\caption{Figure shows how the transition rate for a comoving detector coupled to a field in de Sitter space responds to variations in the energy gap. We include a selection of field temperatures through the choices of $H=1/10$ (green), $H=1$ (blue) and $H=3$ (yellow). The detection time has been taken to be suitably long to avoid transience.}
		\label{fig:RatedScom_omega}
\end{figure}
While in Fig.~\ref{fig:TEDRcomdS}, we have chosen specific values for the energy gap and the Hubble constant, the results apply more generally. Firstly, by increasing the field temperature (via the Hubble constant $H$) as in Fig.~\ref{fig:TEDRcomdS_2}, the detector thermalises more quickly. Conversely, a decrease in field temperature means the transient phase is dominant for longer, extending the time needed to reach thermal equilibrium. Secondly, in Fig.~\ref{fig:RatedScom_omega},  we examine how the transition rate responds to variations in energy gap. The general trend we observe is that the transition rate is high in the region where the energy gap is negative and decreases steadily into the positive energy gap region. This is as expected. Just as it is more likely for an atom in an excited state to emit rather than absorb a photon, a detector which has absorbed a field quanta is more likely to de-excite by emitting a field quanta ($\omega<0$) rather than to absorb ($\omega>0$). The result being that the profiles of Figs.~\ref{fig:RatedScom_tau}--\ref{fig:TEDRcomdS_2} are shifted upwards when the energy gap is negative. 

We further note from Fig.~\ref{fig:RatedScom_omega} that the passage from the negative to the positive energy gap region is sharpest when we take a small value for the Hubble constant, such as $H=1/10$ (green curve), which sees the transition rate approximately vanish in the positive energy gap region. This mirrors the behaviour of an inertial detector in flat space which again, is as expected, as the scale factor $a(t)=e^{Ht}$ `flattens' for small values of the Hubble constant. For larger values of the Hubble constant, the de Sitter  (field) temperature is increased and the non-inertial, curved space correction becomes more dominant. Thus, for $H=1$ (blue) and $H=3$ (yellow), we observe a smoother, more linear transition from the negative to positive energy gap region. Similar behaviour can be found in the case of a stationary black hole where a smoother, more linear profile can be achieved by increasing the temperature of the thermal state beyond the Hawking temperature, see Ref.~\cite{Conroy13}.

\subsection{Geodesic detector}
So far we have considered the simple case of a comoving detector with
$t=\tau$. The more general relation between cosmic time and the proper
time experienced by a detector following a geodesic (but not necessarily
comoving) trajectory in the de Sitter universe is given by
\begin{equation}
	\tau=H^{-1}\text{artanh}\left(\frac{e^{Ht}}{\sqrt{e^{2Ht}+c_{1}^{2}}}\right)+c_2,\quad c_{1}>0,
\end{equation}
where  $c_1$ encodes the choice of geodesic via Eq.~\eqref{geodesicU} and $c_2=-H^{-1}\text{artanh}(1/\sqrt{1+c_{1}^{2}})$ is an integration constant which is fixed to ensure that $t(0)=0$. We use the above relation to solve for $e^{Ht}$ which, following a number
of standard trigonometrical machinations, gives the scale factor wholly
in terms of proper time
\begin{equation}
a(\tau)=\cosh(H\tau)+\sqrt{1+c_{1}^{2}}\sinh(H\tau).
\end{equation}
We see here that the choice $c_{1}=0$ reduces the scale factor to $a(\tau)=e^{H\tau}$
as in the comoving case. We refrain from reproducing the plots of the previous section here for more general geodesic trajectories as the comoving case captures adequately the characteristics of a geodesic detector in the de Sitter universe. Instead, we move on to an accelerated trajectory in the form of a Kodama detector.

\subsection{Kodama detector}
From Eq.~\eqref{fourvel0}, we can read off the relationship between cosmic time and proper time for a Kodama observer
\begin{equation}
	\tau=\int dt\sqrt{1-H^{2}K^{2}},
\end{equation}
where $K$ is the constant Kodama radius from Eq.~\eqref{traj0}. In de Sitter space, the parameter $H\equiv H(t)$ above is constant so that we can integrate in a straightforward manner, see Ref.~\cite{Acquaviva:2011vq}, to find
\begin{equation}
		\label{dSKodtime}
	\tau=\sqrt{V_{dS}}t,\quad \mbox{where}\quad V_{dS}\equiv 1-H^{2}K^{2}.
\end{equation}
Here, we note that $0<V_{dS}<1$, where the upper bound comes from the minimal Kodama radius $K=0$ while the lower bound corresponds to the cosmological apparent horizon at $K=1/H$. Substitution of Eq.~\eqref{dSKodtime} into $a(t)=e^{Ht}$ yields the scale factor 
\begin{equation}
	a(\tau)=e^{\frac{H\tau}{\sqrt{V_{dS}}}}.
\end{equation}
We use the above to compute the Hubble parameter $H(\tau)\equiv \mathring{a}/a$, from which follows the relations $H(\tau)=H/\sqrt{V_{dS}}$ and  $V(\tau)=1/V_{dS}$. We  then use these to explicitly compute the conformal Kodama trajectories \eqref{traj0} which we find to be
\begin{equation}
	\eta(\tau)=-\frac{1}{H}e^{-\frac{H\tau}{\sqrt{V_{dS}}}},\quad r(\tau)=\frac{\sqrt{1-V_{dS}}}{H}e^{-\frac{H\tau}{\sqrt{V_{dS}}}}.
\end{equation}
The geodesic distance is then
\begin{equation}
	\sigma^{2}(s)=-\frac{4V_{dS}}{H^{2}}\sinh^{2}\left(\frac{Hs}{2\sqrt{V_{dS}}}\right).
\end{equation}

Thus, to compute the transition rate for a Kodama detector in de Sitter space, we simply replace $H\to H/\sqrt{V_{dS}}$ in Eq.~\eqref{FcorrdScom} and evaluate to obtain
\begin{equation}
	\label{dS_Kodama_Rate}
	{\cal \dot{F}}_{\tau}(\omega)\equiv\frac{\omega}{2\pi}\left[\frac{1}{e^{\omega/T_{dS}^{\text{loc}}}-1}\right]+J_{\tau},\quad T_{dS}^{\mathrm{loc}}\equiv\frac{H}{2\pi\sqrt{V_{dS}}}.
\end{equation}
We see here that the detailed balance form of the KMS condition is satisfied in the limit of large detection time, with the detector thermalising to the temperature $T_{ds}^{\mathrm{loc}}$. We interpret this temperature as the locally-measured de Sitter temperature whereby the de Sitter temperature is shifted by the factor $\sqrt{V_{dS}}$. As $V_{dS}\equiv 1-H^{2}K^{2}$, the temperature $T_{ds}^{\mathrm{loc}}$ blows up on the cosmological apparent horizon, i.e. when the Kodama radius $K\to1/H$, and reduces to a comoving geodesic trajectory with $T_{ds}^{\mathrm{loc}}=T_{ds}$ when $K=0$.

	\begin{figure}[htb]
	\centering
	\includegraphics[width=\linewidth]{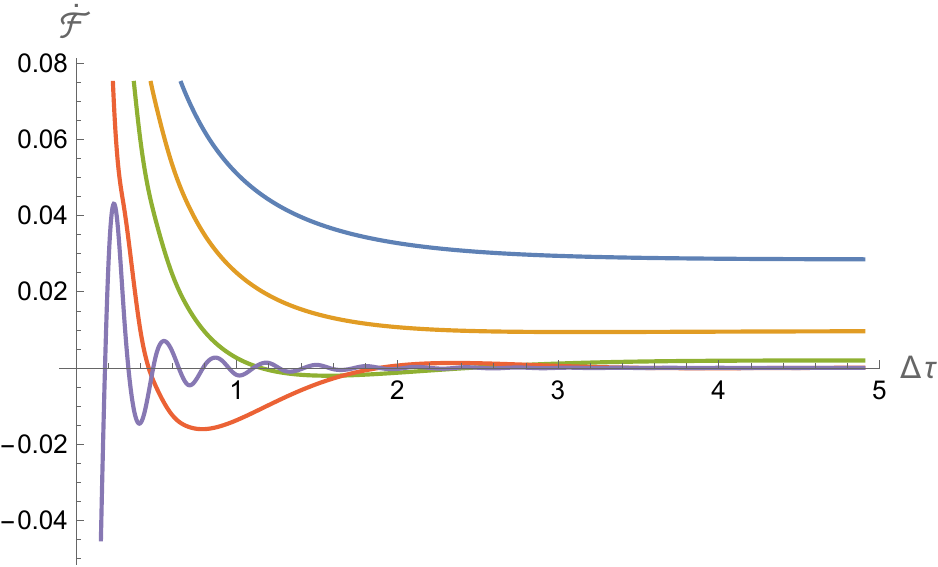}
	\caption{Figure shows the transition rate of a Kodama detector in de Sitter space as a function of detection time. A selection of energy gaps are depicted, i.e. $\omega=1/10$ (blue), $\omega=1/2$ (yellow), $\omega=1$ (green), $\omega=2$ (red), and $\omega=20$ (purple). In each case, we have set the field temperature to $T=1/\pi$ by choosing $H=2$.}
	\label{fig:Kodama_dS_1}
\end{figure} 
In Fig.~\ref{fig:Kodama_dS_1}, we plot the transition rate for a Kodama detector as a function of detection time. The results follow the same overall trend as in the comoving case, cf. Fig.~\ref{fig:RatedScom_tau}, with transient effects diminishing over time and the oscillatory nature of these effects becoming more apparent when the energy gap is large.

	\begin{figure}[htb]
	\centering
	\includegraphics[width=\linewidth]{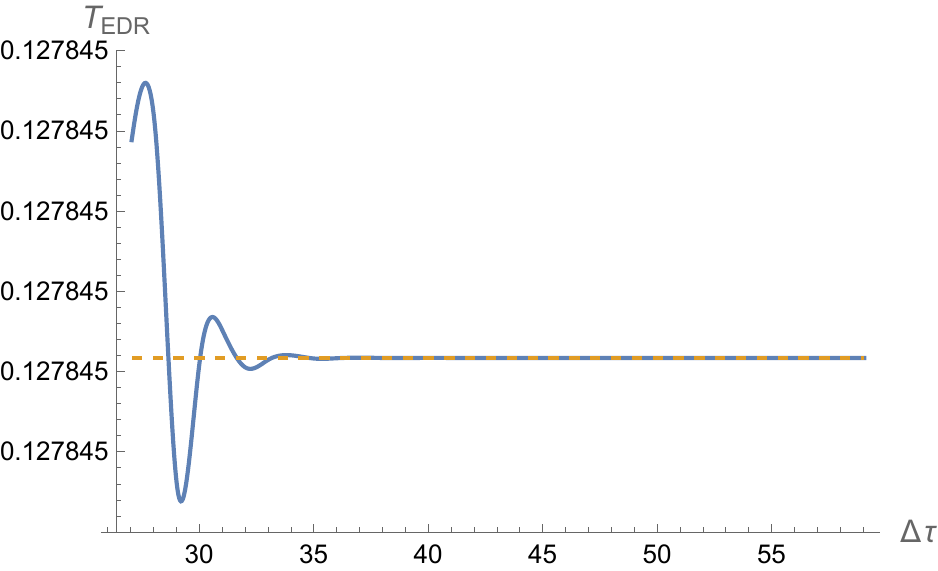}
	\caption{Figure shows the temperature estimator $T_{EDR}$ defined in Eq.~\eqref{TEDR} as a function of detection time for a Kodama detector in the de Sitter universe. Early transient oscillations give way to a constant value which coincides with the locally-measured de Sitter temperature $T_{dS}^{\mathrm{loc}}$, depicted here by the yellow, dashed line. Here we have chosen $H$ and $V_{dS}$ to ensure $T_{dS}^{\mathrm{loc}}\approx 0.12784$ while $\omega=2$.}
	\label{fig:Kodama_dS_2}
\end{figure}

We observe from Fig.~\ref{fig:Kodama_dS_2} that the temperature estimator $T_{EDR}$ for a Kodama detector also has a similar profile to the comoving case, with early transient oscillations giving way to a constant value after a suitably-long detection time which we interpret as the temperature the detector has thermalised to. For the chosen parameters, the detector appears to have approximately thermalised  to the locally-measured de Sitter temperature after a detection time of approximately $\Delta \tau\approx40$. In general, the time taken to approximate thermality is highly dependent on the choice of field temperature which we have chosen to be $T\approx 0.12784$ for reasons that will become clear later on. In the meantime, it is important to define precisely what constitutes a detector that is `sufficiently thermal'. We do so by setting a tolerance of $10^{-10}$ away from the  locally-measured de Sitter temperature. Indeed, the error at $\Delta\approx40$ is of this order so that, for the parameters given in Fig.~\ref{fig:Kodama_dS_2}, we consider a detector to be sufficiently thermalised after this time.


	\begin{figure}[htb]
	\centering
\includegraphics[width=\linewidth]{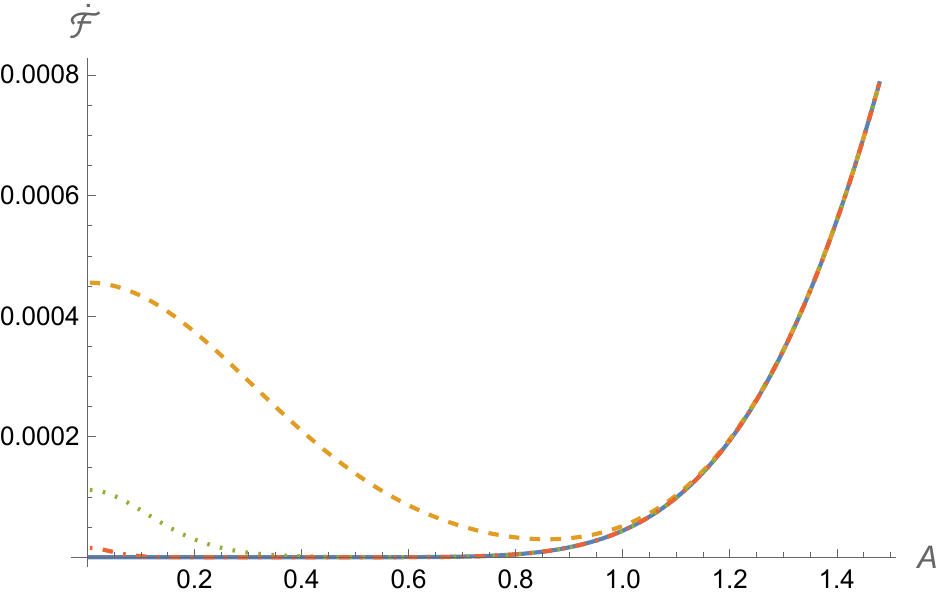}
	\caption{Figure shows the transition rate of a Kodama detector coupled to a field in de Sitter space as a function of acceleration. We observe the expected monotonic increase when the detection time is very large. Here, we increase the detection time from $\Delta\tau=6$ (yellow) to $\Delta\tau=15$ (green) and $\Delta\tau=40$ (red), with the blue curve representing a numerically large detection time. In each case, we have chosen $\omega=2$ and $V_{dS}=1/2$.}
	\label{fig:Kodama_dS_3}
\end{figure} 
	\begin{figure}[htb]
\centering
\includegraphics[width=\linewidth]{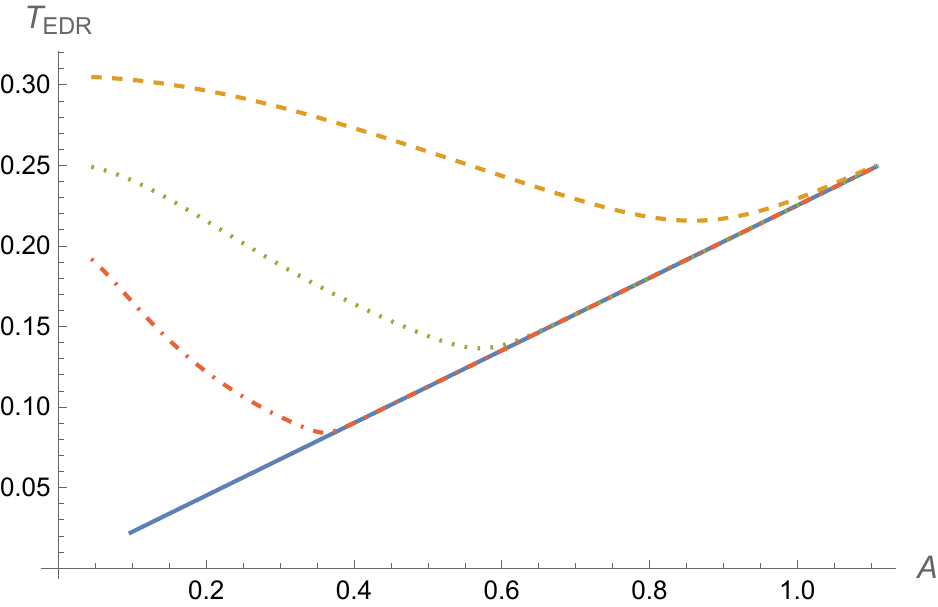}
\caption{Here we investigate the strong anti-Unruh effect by considering the temperature estimator $T_{EDR}$ as a function of acceleration. All parameters have been chosen as in Fig.~\ref{fig:Kodama_dS_3} }
\label{fig:Kodama_dS_4}
\end{figure} 
As the Kodama trajectory is an accelerated one, we can investigate how the detector responds to increased acceleration. The Kodama detector we are considering is accelerating with a magnitude of 
\begin{equation}
	\label{dSacc}
	A=\biggl|\frac{KH^2}{\sqrt{1-K^2 H^2}}\biggr|=2\pi \sqrt{1-V_{dS}} T_{dS}^\mathrm{loc},
\end{equation}
which blows up when the Kodama radius $K$ approaches the horizon $1/H$.

 The expectation from the Unruh effect is that both the transition rate and the temperature estimator $T_{EDR}$ of a detector which has thermalised (or approximately thermalised) will increase as the acceleration increases, that is, in accordance with the mantra `accelerating detectors get hotter'. There is some tentative evidence, however, purporting  the existence of an `anti-Unruh effect' whereby a detector may cool down in some region of the parameter space even as the acceleration increases, see Refs.~\cite{Brenna2016,Mann2020,Conroy13,Garay2016,Campos2021}. This has been observed for finite detection times and accelerations of small magnitude in lower-dimensional spacetimes such as a BTZ black hole \cite{Brenna2016,Garay2016} and more, recently in the near horizon throat of an extremal four-dimensional black hole \cite{Conroy13}.  The latter result, while robust, has the caveat that the detector could not be said to have approximated thermal equilibrium and so falls short of direct four-dimensional evidence of this proposed effect. 

We call an anti-correlation between acceleration and transition rate the \emph{weak anti-Unruh effect} while an anti-correlation between acceleration and $T_{EDR}$ is known as the \emph{strong anti-Unruh effect.} We look for evidence of these in Figs.~\ref{fig:Kodama_dS_3} and \ref{fig:Kodama_dS_4} respectively. Beginning with Fig.~\ref{fig:Kodama_dS_3}, we observe clear dips in the profile of the curve as the acceleration increases for detection times of $\Delta\tau=6$ (yellow), $\Delta\tau=15$ (green), and $\Delta\tau=40$ (red) which warrant further investigation. The question we wish to answer is whether this decrease is due to transience or a result of the proposed (weak) anti-Unruh effect. 

What may not be immediately clear from Fig.~\ref{fig:Kodama_dS_3} is that the transition rate for the green curve is still decreasing when the acceleration $A\approx 0.568$. It is this value for the acceleration that leads to the choice of field temperature in  Fig.~\ref{fig:Kodama_dS_2} via the formula given by Eq.~\eqref{dSacc}. This allows us to easily compare the green curve with Fig.~\ref{fig:Kodama_dS_2} and observe that the detector is still dominated by transience in the region where this dip occurs. What then of the red curve with $\Delta\tau=40$ which still appears to decrease with acceleration in some region of the parameter space? On closer inspection this region is confined to accelerations of $A\lessapprox 0.36$ so that the temperature in Fig.~\ref{fig:Kodama_dS_2} must be lowered resulting in a longer period of transience. Indeed, by our definition, such a detector will not become sufficiently thermalised until after a detection time of $\Delta\approx 78$ has passed. As such, we conclude that these dips in transition rate are a result of transience and not evidence of the anti-Unruh effect.

We find a similar story in Fig.~\ref{fig:Kodama_dS_4} where we plot the temperature estimator $T_{EDR}$ as a function of acceleration. The expectation from the Unruh effect is that the temperature of the detector, modelled by $T_{EDR}$, will monotonically increase with acceleration. This monotonic increase is present when the detection time is suitably large (see blue curve) but it is not difficult to find instances where the profile decreases for shorter detection times. Here it is more apparent that this dip is present in the green curve for accelerations $A\lessapprox0.568$ and the red curve for $A\lessapprox 0.36$ but for the reasons given above we put this down to transience. 

Having now introduced the concepts at play and the methodologies at our disposal for analysing the thermal behaviour of both a geodesic and accelerating detector in the context of a well-studied spacetime, we now broaden our investigation to less-restrictive FLRW spacetimes to better understand how a detector responds in an expanding universe. A major advantage of exploring beyond de Sitter space is that while the competing concepts of temperature in dynamical spacetimes detailed in Section~\ref{sec:SurfaceGravity} coincide in de Sitter space, differences can be highlighted in less-restrictive cosmologies. 
\section{A detector in the FLRW universe}
\label{sec:FLRW}
\subsection{A class of tractable solutions}
Recall that in order to transform the metric~\eqref{FLRW} into conformal coordinates, we defined a new time parameter via $dt=ad\eta$. Using this along with the conformal time coordinate for a Kodama trajectory given in Eq.~\eqref{traj0} allows us to write 
\begin{equation}
	\label{timetime}
	t=\int \sqrt{V(\tau)}d\tau,\quad \mbox{where}\quad V(\tau)\equiv1+K^2 H^2(\tau).
\end{equation}
This encodes how cosmic time $t$ is related to the proper time $\tau$ experienced by the Kodama observer in a generic FLRW spacetime.  Much of the difficulty in finding tractable solutions in this context centres on (i) using the above relation to express a scale factor $a(t)$ wholly in terms of proper time $\tau$ and subsequently (ii) integrating Eq.~\eqref{traj0} to find an explicit conformal time trajectory also in terms of $\tau$. These trajectories, along with the scale factor $a(\tau)$, must be known explicitly to make progress in terms of computing the transition rate~\eqref{transitionsharp}.

To this end, let us consider a class of solutions defined by the Ansatz
\begin{equation}
	\label{Ansatz}
	\frac{d\eta}{d\tau}=a^n,
\end{equation}
for some integer $n$. It is clear from Eq.~\eqref{traj0} that $d\eta/d\tau=\sqrt{V(\tau)}/a$ which allows us to express the above Ansatz as the differential equation
\begin{equation}
	\label{diff}
	a^2+K^2 \mathring{a}^2=a^{2(n+2)},
\end{equation}
where the superscript $\mathring{~}$ indicates a derivative w.r.t. proper time $\tau$, i.e. $\mathring{a}=da/d\tau$.  The reason for choosing such an Ansatz is that we may use the above differential equation to solve for a scale factor $a(\tau)$. As long as this scale factor (and its powers) are analytically integrable, we can use Eq.~\eqref{traj0} to find an expression for the conformal time trajectory, before using Eq.~\eqref{timetime} to return the scale factor to cosmic time $t$. Having the scale factor in terms of cosmic time as well as proper time allows for both a global interpretation of the cosmology and the local instantaneous measurements required for the particle detector approach. 

To solve Eq.~\eqref{diff}, we rewrite like so
\begin{equation}
	\frac{d\tau}{K}=\pm\frac{da}{a\sqrt{a^{2(n+1)}-1}},
\end{equation}
and integrate to find
\begin{equation}
	\label{deltau1}
	\tau=\pm\frac{K\arctan\sqrt{a^{2(n+1)}-1}}{n+1},
\end{equation}
up to an integration constant. Solving for $a$ then gives
\begin{equation}
	\label{atau}
	a(\tau)=\sec\left(\frac{(n+1)\tau}{K}\right)^{\frac{1}{(n+1)}},
\end{equation}
which is valid on the interval $-\frac{\pi K}{2(n+1)}\leq \tau<\frac{\pi K}{2(n+1)}$ and we have chosen the integration constant to ensure that $a(0)=1$. With the scale factor in this form, one can use Eq.~\eqref{timetime} to solve for cosmic time
\begin{equation}
	t=-\frac{K}{n+1}\artanh\sin\left(\frac{(n+1)\tau}{K}\right),
\end{equation}
which, in turn, implies
\begin{equation}
	\tau=-\frac{K}{n+1}\arcsin\tanh\left(\frac{(n+1)t}{K}\right).
\end{equation}
Substituting the above expression for $\tau$ into Eq.~\eqref{atau} gives the scale factor in terms of cosmic time $t$,
\begin{equation}
	\label{at}
	a(t)=\text{\ensuremath{\cosh}}^{n+1}\left(\frac{(n+1)t}{K}\right).
\end{equation}

We have now arrived at a class of scale factors which are tractable within the Kodama prescription in that the integral in the conformal time coordinate~\eqref{traj0} can be evaluated analytically. We have both the explicit expression for the scale factor in terms of proper time~\eqref{atau} required to evaluate the transition rate~\eqref{transitionsharp}, as well as the above expression in terms of cosmic time so that we can understand the global properties of the spacetime. 

Interestingly, the class of scale factors we have derived describe a non-singular, bouncing cosmology in that they satisfy the criteria of (i) being non-degenerate and positive for all $-\infty <t<\infty$,  and (ii) they satisfy the bounce conditions
\begin{equation}
	H\rvert_{t=0}=0,\qquad \dot H\rvert_{t=0}>0.
\end{equation}
As $\cosh$ is an even function, Eq~\eqref{at} satisfies these conditions for all integers $n\geq0$. Furthermore, and perhaps more relevantly for this investigation, the scale factor \eqref{at} asymptotes to de Sitter space at late times which aids comparison with the previous section. 
\subsection{Geodesic detector}
We look now at a geodesic detector in an FLRW universe with a scale factor of the form
\begin{equation}
	a(t)=\cosh(\lambda t).
\end{equation}
This is part of the class of solutions identified in Eq.~\eqref{at}, where we have chosen $n=0$ and it is understood that the dimensionful parameter $\lambda$ is proportional to field temperature\footnote{As this scale factor asymptotes to de Sitter space, if we choose $\lambda=H_{dS}/\sqrt{V_{dS}}$, the detector will thermalise to the locally-measured de Sitter temperature $T_{dS}^\mathrm{loc}$ at the limit of large detection time, as does a Kodama detector in de Sitter space. For a geodesic and comoving detector, choosing $\lambda=H_{dS}$ results in the detector thermalising to $T_{dS}$ as in Eq.~\eqref{RatedScominf}. In these instances, $H_{dS}$ is the Hubble constant.} $T_{RW}\equiv\lambda/2\pi$. The relationship between cosmic time and proper time for a geodesic detector in this spacetime follows from the time component of Eq.~\eqref{geodesicU}, which yields
\begin{equation}
t =\lambda^{-1}\text{arsinh}\left(\sqrt{c_{1}^{2}+1}\sinh(\lambda\tau)\right).
\end{equation}
We see here that the choice of $c_1=0$, which corresponds to a comoving detector, produces the relation to $t=\tau$ as required. With this in hand, we find the proper time scale factor to be
\begin{equation}
	\label{ataugeodesic}
	a(\tau)=\sqrt{\cosh^2(\lambda\tau)+c_1^2 \sinh^2(\lambda \tau)},
\end{equation}
and the non-vanishing trajectories in conformal coordinates \eqref{geotrajc} to be
\begin{align}
	\eta(\tau)&=\lambda^{-1}\arctan\biggl(\sqrt{1+c_1^2}\sinh(\lambda\tau)\biggr),
	\nn\\ r(\tau)&=\lambda^{-1}\arctan\biggl(c_1 \tanh(\lambda\tau)\biggr).
	\label{FLRWtrajgeo}
\end{align}

	\begin{figure}[htp]
	\centering
	\includegraphics[width=\linewidth]{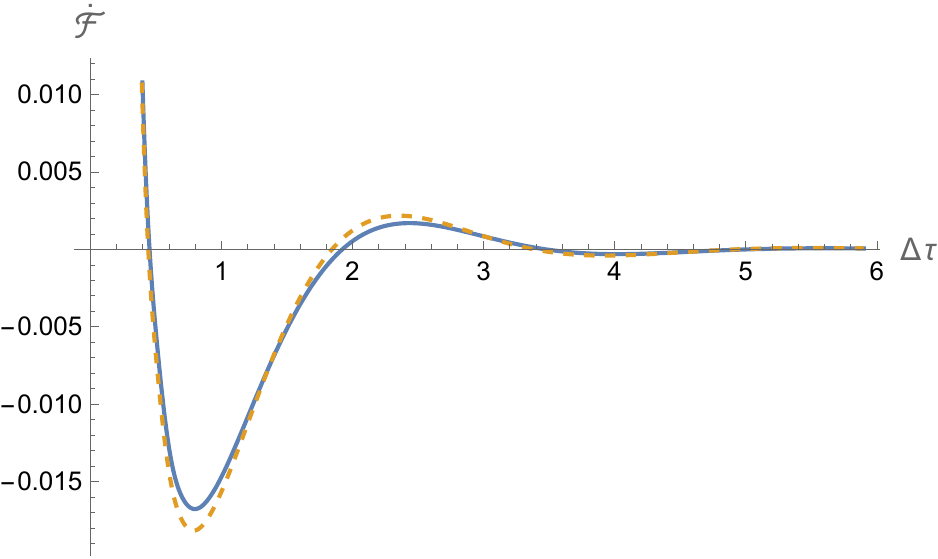}
	\caption{Plot demonstrating the weak dependence of a geodesic detector on the parameter $c_1$ in the trajectories \eqref{FLRWtrajgeo} which encodes the choice of geodesic. The blue curve corresponds to a comoving detector with $c_1=0$ which displays only minor differences in profile to the choice of $c_1=5000$ (yellow curve).} 
	\label{fig:FLRW_geodesic_compare}
\end{figure}

We now have all the necessary components of the geodesic distance $\sigma^2(\tau,s)$ to compute the sharp-switching transition rate \eqref{TranRateSharp}. The first thing to note is that the transition rate for a geodesic detector is only very weakly dependent on the choice of geodesic via the parameter $c_1$. We see this in Fig.~\ref{fig:FLRW_geodesic_compare} where a dramatic increase in $c_1$ to $c_1=5000$ (yellow) results in only a superficial change in profile from the comoving case $c_1=0$ (blue). For later detection times, plotting transition rate as a function of $c_1$ gives a constant profile. As such, it is reasonable to consider only the simple case of a comoving detector with $c_1=0$. In this case $t=\tau$ so that the scale factor \eqref{ataugeodesic} becomes $a(\tau)=\cosh{(\lambda \tau)}$, yielding the trajectories
\begin{equation}
	\eta(\tau)=\lambda^{-1}\arctan \sinh(\lambda \tau),\quad r(\tau)=r_0.
\end{equation}

		\begin{figure}[htp]
	\centering
\includegraphics[width=\linewidth]{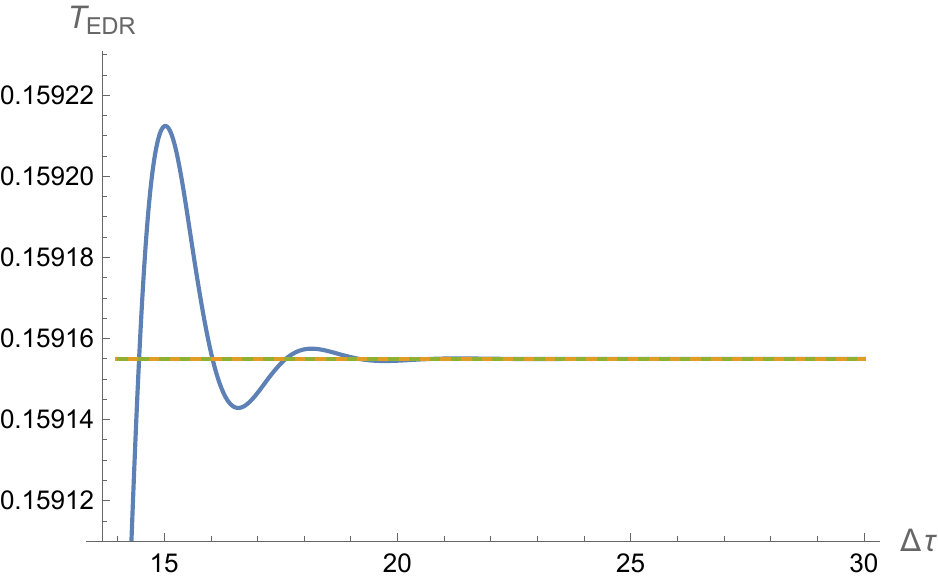}
	\caption{Figure shows how the temperature estimator $T_{EDR}$ (blue curve) for a comoving detector coupled to a field in an FLRW universe with $a(t)=\cosh(\lambda t)$ develops over detection time. Transient oscillations give way to a constant value which coincides with both the Hayward-Kodama temperature $T_{HK}$ (yellow) and the effective temperature $T_{eff}$ (green), which are indistinguishable for the given parameters. Here, we have chosen $\omega=2$ and $\lambda=1$.}
	\label{fig:FLRW_TEDR_Comoving_1}
\end{figure}
In Fig.~\ref{fig:FLRW_TEDR_Comoving_1}, we plot the temperature of a comoving detector via $T_{EDR}$ as a function of detection time. We see here that once the detection time is suitably long so as to avoid transient effects, $T_{EDR}$ becomes approximately constant and aligns with both the Hayward-Kodama temperature $T_{HK}$ given in yellow and the effective temperature $T_{eff}$ given in green. These temperatures are indistinguishable for the given parameters which have been chosen for comparison with Fig.~\ref{fig:TEDRcomdS}. Indeed, the correction coming from the amended spacetime is largely negligible for detection times in this range so that there is little discernible difference between the profiles of Fig.~\ref{fig:TEDRcomdS} and Fig.~\ref{fig:FLRW_TEDR_Comoving_1}. To highlight the differences between $T_{HK}$ and $T_{eff}$, we must tweak the parameters so that transient effects are distilled earlier when the detection time is short, i.e. when the spacetime is most distinct from de Sitter space.

\begin{figure}[htp]
	\centering
	\includegraphics[width=\linewidth]{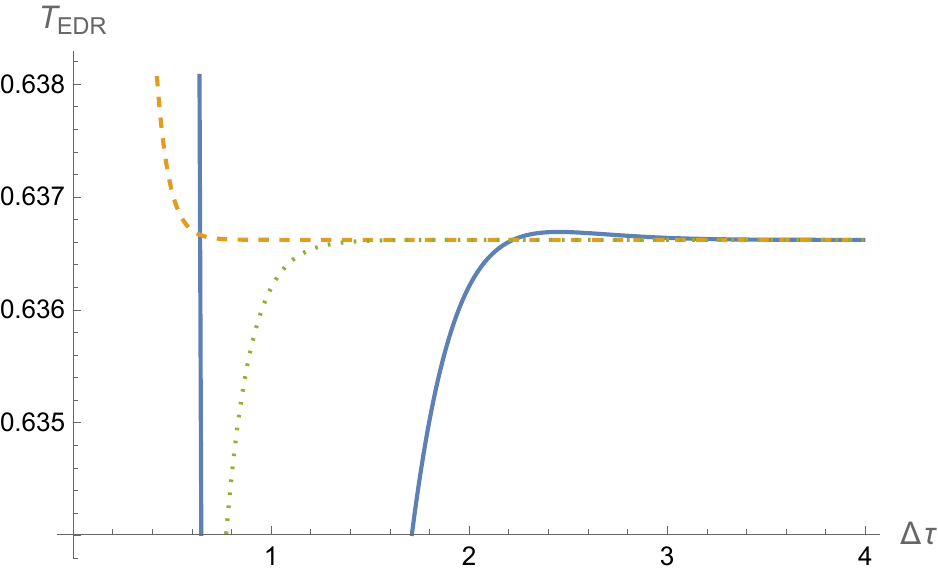}
	\caption{As in Fig.~\ref{fig:FLRW_TEDR_Comoving_1}, figure shows $T_{EDR}$ (blue curve) as a function of time for a comoving detector with $a(t)=\cosh(\lambda t)$. Here we raise the field temperature by increasing $\lambda\to 4$.}
	\label{fig:FLRW_TEDR_Comoving_2}
\end{figure}
To do this, we raise the field temperature via the parameter $\lambda$ where in Fig.~\ref{fig:FLRW_TEDR_Comoving_2} we choose $\lambda=4$. Raising the temperature of the field increases the temperature of the detector so that transient oscillations become sub-dominant earlier. Now, we can see clear differences between $T_{HK}$ (yellow) and $T_{eff}$ emerging for short detection times while both approaches equally yield the thermalisation temperature $T_{RW}\equiv \lambda/(2\pi)$, and consequently the temperature of the dynamical horizon,  so long as the detection time is suitably long. For the given parameters, both approaches asymptote to $T_{RW}$ while the detector is still dominated by transience. 

	\begin{figure}[!htp]
	\centering
	\subfloat[$\omega=0.1$]{
		\includegraphics[width=\linewidth]{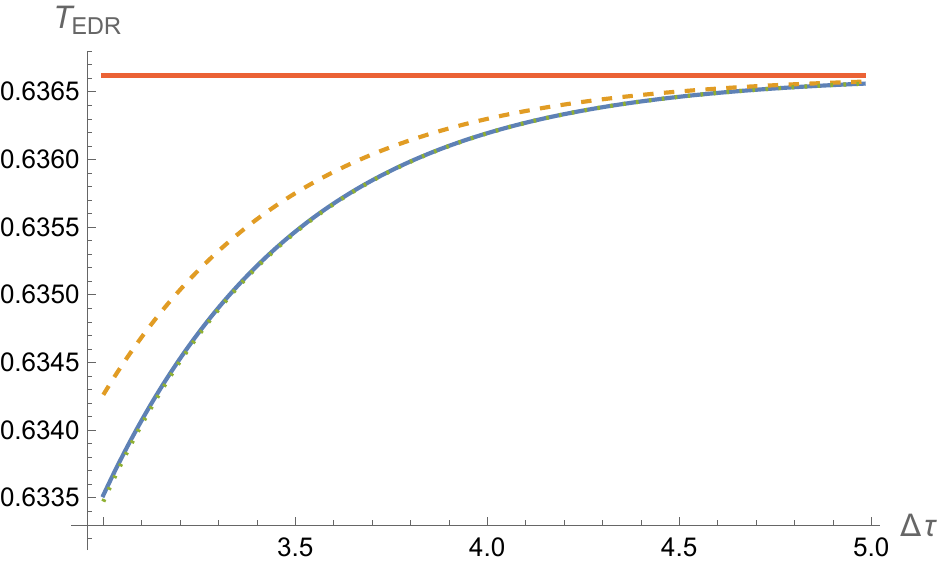}
	}\\
	\subfloat[$\omega=1$]{
	\includegraphics[width=\linewidth]{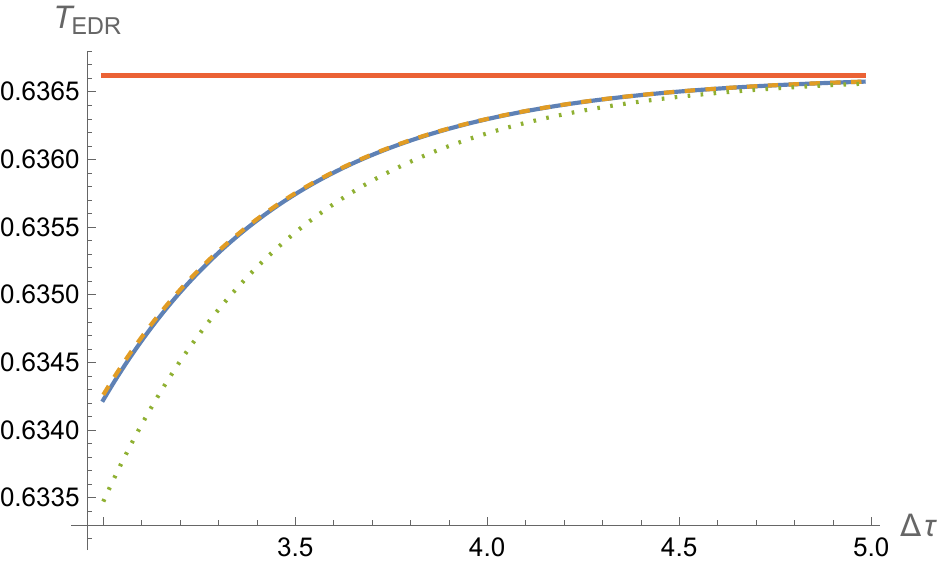}
	}\\
	\subfloat[$\omega=2$]{
		\includegraphics[width=\linewidth]{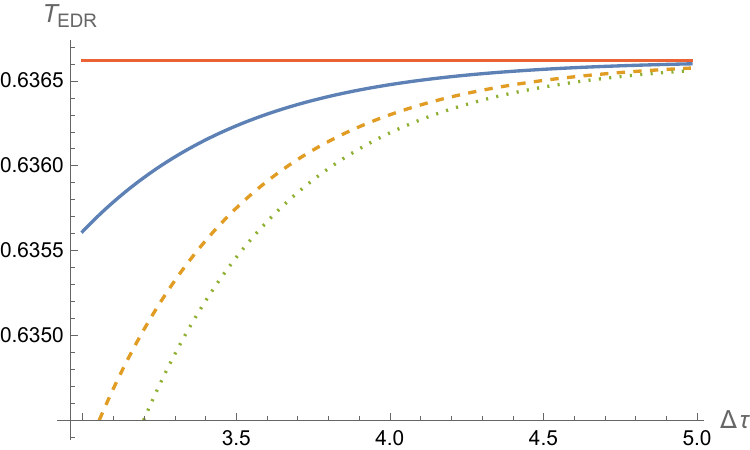}
	}
	\caption{In each case, the blue curve shows how the temperature estimator $T_{EDR}$ of a comoving detector coupled to a field in an FLRW spacetime with $a(t)=\cosh^4(\lambda t/4)$ develops over (detection) time. Alongside this, we depict the Hayward-Kodama temperature $T_{HK}$ (yellow), the effective temperature $T_{eff}$ (green), and the field temperature $T_{RW}$ (red) for a selection of energy gaps where, in each case, we have chosen $\lambda=4$.}
	\label{fig:FLRW_Geodesic_Cosh4}
\end{figure}
In Fig.~\ref{fig:FLRW_Geodesic_Cosh4}, we tweak the scale factor Eq.~\eqref{at} by choosing $n=3$ to give $a(t)=\cosh^4(\lambda t/4)$ and plot $T_{EDR}$ over detection time for a selection of energy gaps. Again, we see clearly that the detector thermalises to the appropriate field temperature when the detection time is suitably long, as do both  $T_{HK}$ and $T_{eff}$. In the region of the parameter space depicted, the temperature of the detector is still dependent on the energy gap in that the slowly increasing profile of Fig.~\ref{fig:FLRW_Geodesic_Cosh4}(a) asymptotes more readily to $T_{RW}$ when the energy gap is increased as in Fig.~\ref{fig:FLRW_Geodesic_Cosh4}(c). This is consistent with prior results regarding thermalisation at the limit of large energy gap, see Refs.~\cite{Conroy13,HodgkinsonLouko2012}.

		\subsection{Kodama dectector}
	Briefly, we turn our attention to a Kodama detector accelerating through an FLRW universe with scale factor \eqref{at}. We choose $n=0$ so that $a(t)=\cosh(t/K)$, where $K$ is the (constant) Kodama radius. From Eq.~\eqref{atau} we have $a(\tau)=\sec(\tau/K)$ which places an upper bound of $\tau<K \pi/2$ on proper time, while the trajectories
		\begin{equation}
			\eta(\tau)=\tau,\quad r(\tau)=K \cos(\tau/K),
		\end{equation}
	follow from Eq.~\eqref{traj0}. The magnitude of the acceleration is given by
	\begin{equation}
		A=\biggl|\frac{K\dot H+K H^2(1-H^2K^2)}{(1-H^2K^2)^{3/2}}\biggr|,
	\end{equation}
which reduces to Eq.~\eqref{dSacc} when the Hubble parameter is constant \cite{Acquaviva:2011vq}.
	
				\begin{figure}[htp]
		\centering
		\includegraphics[width=\linewidth]{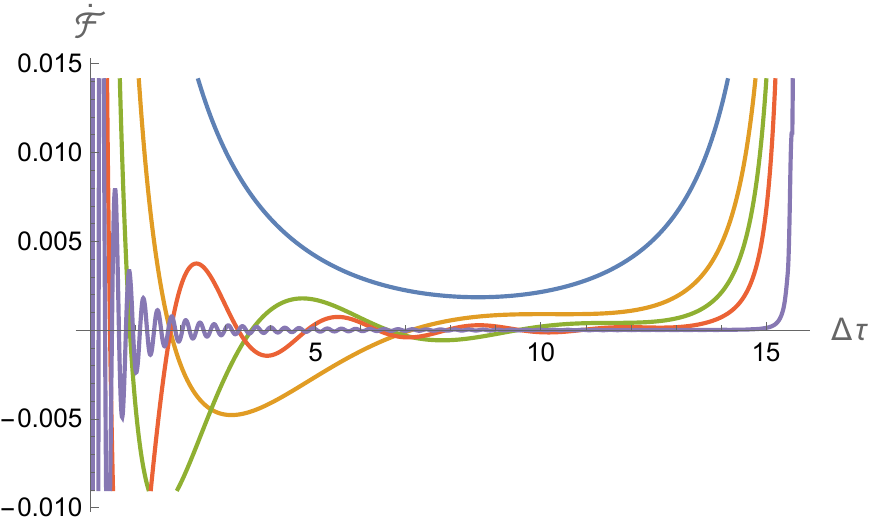}
		\caption{Figure shows the transition rate as a function of detection time for a Kodama detector in the FLRW universe with $a(t)=\cosh(t/K)$. Here we have chosen a Kodama radius of $K=10$ for a selection of energy gaps, i.e. $\omega=1/10$ (blue), $\omega=1/2$ (yellow), $\omega=1$ (green), $\omega=2$ (red), and $\omega=20$ (purple). The transition rate is dominated by noise at both extremes, from the transience at short detection times to the divergence at later times. }
		\label{fig:FLRW_Kodama_1}
	\end{figure}
In Fig.~\ref{fig:FLRW_Kodama_1}, we plot the transition rate as a function of detection time just as in Fig.~\ref{fig:RatedScom_tau} where, in that case, we considered a comoving detector in de Sitter space. As with Fig.~\ref{fig:RatedScom_tau}, we see that transient oscillations dominate when the detection time is short. While these oscillations are damped over time, here we observe an additional divergence as the detection time approaches its upper bound. Our interpretation of Fig.~\ref{fig:FLRW_Kodama_1} is that it depicts a detector which never reaches thermal equilibrium, dominated as it is by noise for both short and long detection times. Due to this noise, $T_{EDR}$ cannot be said to be a good estimator of the detector's temperature and indeed the excitation to de-excitation ratio periodically dips below zero meaning that Eq.~\eqref{TEDR} is undefined in regions of the parameter space. 

		\begin{figure}[htp]
	\centering
	\includegraphics[width=\linewidth]{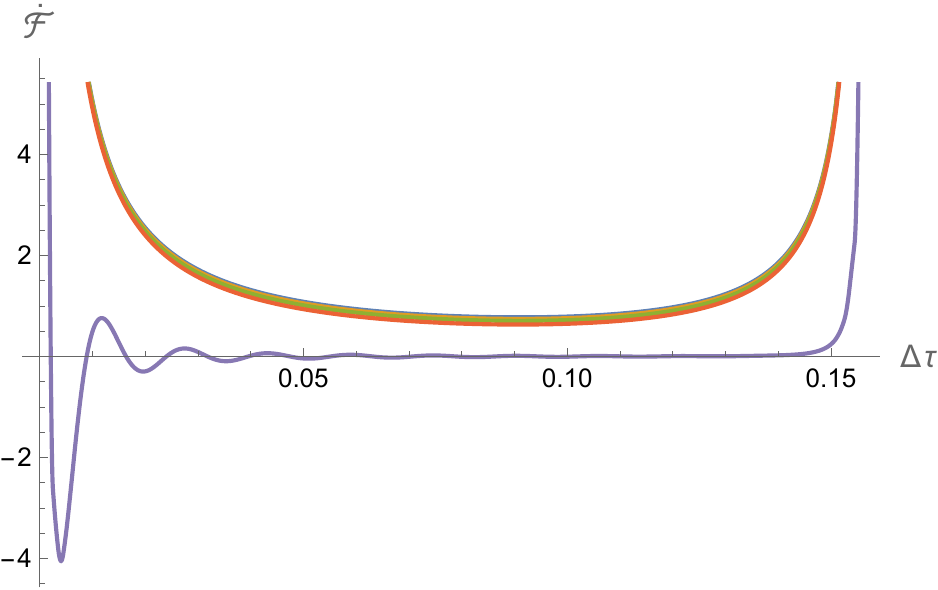}
	\caption{We increase the field temperature by decreasing the areal radius $K$ to $K=1/10$. The selection of energy gaps are as in Fig.~\ref{fig:FLRW_Kodama_1} with the exception of the purple curve which has been increased to $\omega=1000$ to demonstrate the oscillatory nature of the curve.}
	\label{fig:FLRW_Kodama_2}
	\end{figure}
An increase in field temperature could in theory allow the detector to thermalise more quickly and to dominate over the noise as we saw in previous examples. We increase the field temperature via the parameter $K$ in Fig.~\ref{fig:FLRW_Kodama_2}, where field temperature $T\propto 1/K$. The parameter $K$, however, also corresponds to the areal radius and controls the upper bound of the detection time. As such, we can not independently raise the field temperature so that what we see in Fig.~\ref{fig:FLRW_Kodama_2} is merely a curtailed version of Fig.~\ref{fig:FLRW_Kodama_1}.  For lower energy gaps, the curves of Fig.~\ref{fig:FLRW_Kodama_2} have a similar profile to the blue curve of Fig.~\ref{fig:FLRW_Kodama_1}, while increasing the energy gap further gives the distinctive oscillatory behaviour depicted by the purple curves. As a result, we find no region of the parameter space whereby $T_{EDR}$ can be considered a good estimator of the temperature of the detector.

\section{Discussion}
\label{sec:Discussion}
In this paper, we investigated the response of a  particle detector coupled to a field in the conformal vacuum of an FLRW universe. Working within the Unruh-DeWitt particle detector model, we considered an idealised two-level quantum mechanical measuring device, or detector, as it travels along a given trajectory through the expanding universes of de Sitter space and an FLRW spacetime with a scale factor of the form $a(t)\sim \cosh(\lambda t)$. The latter is of particular interest because it is tractable in the sense that the Kodama trajectories can be derived explicitly, as shown in Section \ref{sec:FLRW}, and it asymptotes to de Sitter space for large cosmic time.

Regarding a detector in the de Sitter universe, we gave an overview of how the transition rate, or rate of particle detection, develops as a function of detection time for a detector travelling along a comoving, geodesic, and (accelerated) Kodama trajectory. While de Sitter space is a particularly well-studied spacetime and key results such as Eq.~\eqref{RatedScominf} may be well-known, much of the literature in the field of Unruh-DeWitt detectors works within the context of the de Sitter-Schwarzschild metric, see for example Refs.~\cite{Ng2014,HodgkinsonLoukoOttewill}. As such, we believe it instructive to give explicit details of these results in the context of the de Sitter universe in FLRW coordinates, where we consider a detector coupled to a field in the so-called `conformal vacuum'. Indeed, our analysis goes somewhat beyond what can be found in the literature, for example Ref.~\cite{Acquaviva:2011vq,Ali:2020gij}, which gives an analysis based on the Schlicht  regularisation introduced in Ref.~\cite{Schlicht2004}, particularly with regards to the discussion surrounding the proposed anti-Unruh effect for an accelerating Kodama trajectory. We found that while it is not difficult to find regions of the parameter space where the detector appears to cool with increased acceleration, these are generically found only in regions dominated by transience, meaning that direct evidence of the anti-Unruh effect in a four-dimensional spacetime remains elusive. This is in keeping with related results in flat space, Ref. \cite{Dappiaggi2021}.

Particular attention in the article is paid to the temperature of the detector as modelled through the parameter $T_{EDR}$ as a means of investigating surface gravity and temperature in spacetimes with an evolving horizon. We discussed the concept of surface gravity on dynamical horizons in Section \ref{sec:SurfaceGravity} by introducing various definitions which persist in the literature such as Hayward-Kodama surface gravity; the effective approach of Ashtekar et al.; an amended form of Fodor surface gravity; and the approach of Nielsen-Visser. The latter three were originally conceived in the context of a dynamical black hole and become indistinguishable in an FLRW universe when evaluated on the cosmological apparent horizon so that we grouped these together under the parameter $\kappa_{eff}$ (and the associated temperature $T_{eff}$) given by Eq.~\eqref{effectivesurface}. We tracked the temperature of the detector via the temperature estimator $T_{EDR}$ as it thermalised to the associated field temperature, and by extension the temperature on the cosmological horizon, while similarly investigating the approaches to temperature outlined above.

The simplicity of the de Sitter metric allowed us to derive the thermalisation temperature of the detector analytically. We showed this explicitly in Eqs.~\eqref{RatedScominf} and~\eqref{dS_Kodama_Rate} where we found that a comoving detector thermalises to the de Sitter temperature $T_{dS}$, while an accelerated Kodama detector thermalises to the locally-measured de Sitter temperature $T_{dS}^{\mathrm{loc}}$ which has been modulated by a red-shift factor. While for these analytic results we considered the limit of infinite detection time, we also examined how $T_{EDR}$ behaves over finite time scales as in Fig.~\ref{fig:TEDRcomdS}, Fig.~\ref{fig:TEDRcomdS_2}, and Fig.~\ref{fig:Kodama_dS_2}. In de Sitter space, however, the constant nature of the Hubble parameter means that $T_{eff}$ and $T_{HK}$ are interchangeable so that we required a less restrictive spacetime to shine a light on the differences between these approaches.

We proceeded then with a thermal analysis of the detector in an FLRW spacetime governed by a scale factor of the form $a(t)=\cosh(\lambda t)$. As the accelerated Kodama detector is dominated by noise, the most interesting results were found for a comoving detector in this spacetime. By tweaking the parameters, we finally began to discern some differences between $T_{eff}$ and $T_{HK}$ in Fig.~\ref{fig:FLRW_TEDR_Comoving_2} and Fig.~\ref{fig:FLRW_Geodesic_Cosh4}. In the former, raising the field temperature via the parameter $\lambda$ resulted in thermalisation occurring sooner and transience being distilled earlier. In the case of Fig.~\ref{fig:FLRW_TEDR_Comoving_2}, $T_{RW}$ was reached while the detector was still within its transient phase. 

By tweaking the scale factor to $a(t)=\cosh^{4}(\lambda t/4)$ as in Fig.~\ref{fig:FLRW_Geodesic_Cosh4}, the thermal behaviour became more clear. Again, we saw that a comoving detector thermalises to the field temperature $T_{RW}$ when the detection time is suitably long. For the short detection times shown, the temperature estimator $T_{EDR}$  displays some dependence on the energy gap $\omega$ with a more constant profile emerging as the energy gap is increased. This is in accordance with previous reports \cite{Conroy13,HodgkinsonLoukoOttewill} of detectors thermalising at the limit of large energy gap. Moreover, both $T_{HK}$ and $T_{eff}$ asymptote to the required field temperature when the detection time is suitably long meaning both approaches appear to be equally effective in describing the temperature of the evolving horizon.

What then, if anything, is there to discern from these two approaches? In truth, from the vantage point of Fig.~\ref{fig:FLRW_Geodesic_Cosh4} it is difficult to favour one over the other. However, a clue may be found in their formulations. If we look at the definition for Hayward-Kodama surface gravity given by the Eq.~\eqref{kappaH0} one may deduce that the formula $\kappa_{HK}=\tilde r(H^2+\frac{1}{2}\dot{H})$ captures the curvature of the spacetime more completely than the equivalent effective surface gravity formula $\kappa_{eff}=\tilde rH^2$. Indeed, the Hayward-Kodama surface gravity is proportional to the curvature scalar $R=12(H^2+\frac{1}{2}\dot{H})$ in a geometrically-flat FLRW universe. For the spacetimes detailed in this article the $\dot H$ term is dominant at very early times and vanishes at the de Sitter limit where $\kappa_{HK}=\kappa_{eff}$ so that we may tentatively conclude that, in addition to being a more robust top-down formulation of surface gravity in the conceptual sense, $T_{HK}$ captures the dynamics of curvature more effectively. For the asymptotically-dS spacetimes studied here, this is most apparent at early cosmic times and for shorter detection times.

To make further progress and to better understand these approaches in an expanding universe, one path is to extend the study further into the past. As the class of scale factors identified in Section \ref{sec:FLRW} are non-singular, we can, in theory,  extend the scope of our study to the infinite past which may allow us to distil the transient noise associated with turning the detector on sharply. This could, in theory, allow us to clean up our results, particularly in the case of an accelerated Kodama detector, while an  investigation into the response of a detector as it traverses the `bounce point' of a non-singular bouncing cosmology could be interesting in its own right.

	\appendix
	\section{Mode decomposition \& Wightman Function}
	\label{sec:Mode}
	Consider a scalar field propagating through a Friedmann-Lemaître-Robertson-Walker (FLRW) metric 
	\begin{equation}
		\label{FLRWa}
		ds^2=-dt^2+a^2(t)\left(dr^2+r^2d\Omega^2\right),
	\end{equation}
	with spatially-flat geometry. By introducing a new time parameter $\eta$, defined via $d\eta=a^{-1} dt$, we may express the metric in conformal coordinates like so
	\begin{equation}
		\label{FLRWconfapp}
		ds^2=a^2(\eta)\left(-d\eta^2+dr^2+r^2d\Omega^2\right).
	\end{equation}
	In this form we see, explicitly, the conformally-flat nature of the FLRW metric. A natural choice of vacuum state then is one that is conformal to flat space, which we call the \emph{conformal vacuum} (see, for example, Refs.~\cite{birrell1984quantum,mukhanov2007introduction} for a more detailed treatment). To this end, we consider a massless scalar field $\Phi$ with conformal coupling $\xi=1/6$, characterised by the action         
	\begin{equation}
		S=\frac{1}{2}\int d^4x \sqrt{-g}\left(g^{\mu\nu}\partial_\mu\Phi \partial_\nu \Phi -\frac{1}{6} R\Phi^2\right).
		\label{Actionscalar}
	\end{equation}
	By introducing an auxiliary field $\phi\equiv a \Phi$, we may write the field equations in conformal coordinates like so 
	\begin{equation}
		\label{phifieldeq}
		\left(\partial_\eta^2-\partial^2\right)\phi=0,
	\end{equation}
	where $\partial^2=\delta^{ij}\partial_i\partial_j$ is the Laplace operator and $\phi\equiv\phi(x)$ with $x$ accounting for both temporal and spatial coordinates. Due to the choice of coupling, these field equations are invariant under conformal transformations. Transforming the field $\phi$ into Fourier space via
	\begin{equation}
		\label{Fourier}
		\phi(x)=\int \frac{d^3\textbf{k}}{(2\pi)^{3/2}}\phi_\textbf{k}(\eta)e^{i\textbf{k}\cdot \textbf{x}},
	\end{equation}
	allows us to express the field equation~\eqref{phifieldeq} in terms of the time-dependent, real mode $\phi_\textbf{k}\equiv \phi_\textbf{k}(\eta)$ like so
	\begin{equation}
		\phi_\textbf{k}^\ppr+\omega_k^2 \phi_\textbf{k}=0,\quad \omega_k=\rvert k \rvert,
	\end{equation}
	which has the general solution
	\begin{equation}
		\phi_\textbf{k}=\frac{1}{\sqrt{2}}\left(a_\textbf{k}^-v_k^*+a^+_{-\textbf{k}}v_k\right).
	\end{equation}
	Here, $a^\pm$ are complex constants of integration dependent only on the vector $\textbf{k}$ satisfying $a^+_\textbf{k}=(a^-_\textbf{k})^*$, while the mode functions $v_k$ are normalised\footnote{In writing down the normalization condition~\eqref{normalis}, one must take note of the following identities: As  $\phi$ is real, $\phi^*=\phi$; from~\eqref{Fourier}, we have $\phi_\textbf{k}^*=\phi_{-\textbf{k}}$; and as $v_{k}$ depends on $|k|$, we have $v_k=v_{-k}$. This, along with the identity $a^+_\textbf{k}=(a^-_\textbf{k})^*$, gives $a_{\textbf{k}}^{+}a_{\textbf{k}}^{-}=a_{-\textbf{k}}^{-}a_{-\textbf{k}}^{+}$, which is required to write down~\eqref{normalis}.}.  such that 
	\begin{equation}
		\label{normalis}
		\phi_{\textbf{k}}\partial_{\eta}\phi_{\textbf{k}}^{*}-\phi_{\textbf{k}}^{*}\partial_{\eta}\phi_{\textbf{k}}=\frac{1}{2}\biggl[v_{k}^{\prime}v_{k}^{*}-v_{k}v_{k}^{*\prime}\biggr]=i.
	\end{equation}
	This normalisation condition is simply a result of defining the scalar product\footnote{Here, $d\Sigma$ is the volume element of a spacelike hypersurface $\Sigma$ (assumed to be a Cauchy surface) and we define $d\Sigma^\mu\equiv n^\mu d\Sigma$ for a future-directed unit vector $n^\mu$ orthogonal to the hypersurface.}
	\begin{equation}
		(\phi_1,\phi_2)=-i\int_\Sigma d\Sigma^\mu\sqrt{-g_\Sigma}\;\phi_1 \overleftrightarrow{\partial_\mu} \phi_2^*,
	\end{equation}
	and noting that there exists a complete set of mode solutions $v_k$ satisfying
	\begin{equation}
		(v_k,v_{k^\pr})=\delta_{kk^\pr},\quad (v_k^*,v^*_{k^\pr})=-\delta_{kk^\pr},\quad (v_k,v^*_{k^\pr})=0.
	\end{equation}
	Substitution of~\eqref{normalis} into~\eqref{Fourier} allows us to write down the mode decomposition
	\begin{equation}
		\label{phimodedecomphat}
		\hat\phi(x)	=\frac{1}{\sqrt{2}}\int\frac{d^{3}\textbf{k}}{(2\pi)^{3/2}}\left(\hat a_{\textbf{k}}^{-}v_{k}^{*}e^{i\textbf{k}\cdot \textbf{x}}+\hat a_{\textbf{k}}^{+}v_{k}e^{-i\textbf{k}\cdot \textbf{x}}\right),
	\end{equation}
	where the constants $a^\pm$ have been elevated to creation and annihilation operators satisfying the usual commutation relations
	\begin{equation}
		\label{acomms}
		[\hat a_\textbf{k}^-,\hat a_{\textbf{k}^\pr}^+]=\delta_{\textbf{k}\textbf{k}^\pr},\qquad [\hat a_\textbf{k}^-,\hat a_{\textbf{k}^\pr}^-]=[\hat a_\textbf{k}^+,\hat a_{\textbf{k}^\pr}^+]=0,
	\end{equation}
	following canonical quantization, Refs.~\cite{birrell1984quantum,mukhanov2007introduction}. 
	
	Quantization of the field $\hat \phi$ is achieved by postulating the mode expansion~\eqref{phimodedecomphat} along with the commutation relations~\eqref{acomms} and the normalisation condition~\eqref{normalis}, while the  mode functions $v_k$ are specific to the theory. As $v_k\equiv v_k(\eta)$ form a basis of solutions to~\eqref{phifieldeq}, we can find explicit expressions for these by solving the differential equation
	\begin{equation}
		\label{vexplicit}
		v^\ppr_k+k^2v_k=0\qquad\mbox{to find}\qquad v_k=\frac{e^{i|k|\eta}}{\sqrt{|k|}}.
	\end{equation}
	
	Next, we turn our attention to the two-point Wightman function 
	\begin{equation}
		W(x;x^\prime)\equiv \langle 0 |\Phi(x)\Phi(x^\pr)|0\rangle,
	\end{equation}
	which will be instrumental in understanding the interaction of the detector with the quantum field. Substitution of the mode decomposition~\eqref{phimodedecomphat}, while noting that we have defined the field $\hat\Phi$ in terms of the auxiliary field $\hat \phi$ like so $\hat \Phi\equiv a^{-1}\hat\phi$, gives
	\begin{equation}
		W(x;x^\pr)=\langle 0 | \int\frac{d^3\textbf{k}}{(2\pi)^3}\frac{1}{2|k|}\frac{v_k^*(\eta) v_k(\eta^\pr)}{a(\eta)a(\eta^\pr)}e^{i\textbf{k}\cdot(\textbf{x}-\textbf{x}^\pr)}|0\rangle.
	\end{equation}
	Taking the solution \eqref{vexplicit}, we find
	\begin{equation}
		\label{Wightman0}
		W_\epsilon(x;x^\pr)=\int\frac{d^{3}\textbf{k}}{2(2\pi)^{3}}\frac{e^{i|k|(\eta-\eta^{\prime}-i\epsilon)+i\textbf{k}\cdot(\textbf{x}-\textbf{x}^{\prime})}}{a(\eta)a(\eta^{\prime})|k|},
	\end{equation}
	where we have inserted the small parameter $\epsilon>0$ to ensure that the expression is a distribution. Finally, we evaluate the integral to obtain
	\begin{equation}
		\label{WightmanApp}
		W_\epsilon(x;x^\pr)	=\frac{1}{(2\pi)^{2}}\frac{1}{a(\eta)a(\eta^{\prime})}\frac{1}{-|\eta-\eta^{\prime}-i\epsilon|^{2}+|\mathbf{x}-\mathbf{x}^{\prime}|^{2}}.
	\end{equation}
	\section{Regularisation and sharp-switching}
	\label{sec:Sharp}
To examine the response of the detector explicitly, it is convenient to have an explicit regularisation for the Wightman Green function. This is possible so long as the quantum field is in a state that satisfies the Hadamard condition, thereby following the Hadamard singularity structure. In this case the response function is given by
\begin{align}
	\label{responsereg}
	\nn\mathcal{F}(\omega) & =2\int_{-\infty}^{\infty}du\,\chi(u)\int_{0}^{\infty}ds\,\chi(u-s)
	\nn\\&\times\left(\cos\omega s\;W(u,u-s)+\frac{1}{4\pi^{2}s^{2}}\right)
	\nn\\& +\frac{1}{2\pi^{2}}\int_{0}^{\infty}\frac{ds}{s^{2}}\int_{-\infty}^{\infty}du\,\chi(u)\left[\chi(u)-\chi(u-s)\right]
	\nn\\&-\frac{\omega}{4\pi}\int_{-\infty}^{\infty}du\;\chi^{2}(u),
\end{align}
as in Refs.~\cite{Satz2007,LoukoSatz2008}. The quantity $W(u,u-s)$ is the limit of the Wightman Green function for the scalar wave equation with $\epsilon\to 0^{+}$. $W(u,u-s)$ is regular everywhere except at the coincident limit which occurs when $x(u)=x(u-s)$, or equivalently when $s=0$, but this Hadamard singularity is regularised by the counter-term $1/(4\pi^2)$. That is to say, the response function is well-defined and we are free to take the limit $\epsilon\to 0^{+}$ prior to integration. 

To obtain the sharp-switching limit of the response function, we follow the methodology of Satz in Ref. \cite{Satz2007} by considering a switching function of the form
\begin{equation}
	\label{switchdef}
	\chi(u)=h_{1}\left(\frac{u-\tau_{0}+\delta}{\delta}\right)\times h_{2}\left(\frac{-u+\tau+\delta}{\delta}\right),
\end{equation}
for some initial (proper) time $\tau_{0}$ with $\tau>\tau_{0}$. We then stipulate that $\delta>0$, while $h_{1}$ and $h_{2}$ are smooth functions satisfying
\begin{align}
	\label{hprops}
	\nn
	h_{i}(x) & =0,\qquad\text{for}\qquad x\leq0\\
	h_{i}(x) & =1,\qquad\text{for}\qquad x\geq1.
\end{align}
To understand this choice of switching function, we consider a detector initially in an off position, indicating that the detector will not register any interaction with the quantum field and no particles will be detected. The detector will begin to register interaction as it is switched on smoothly via $h_{1}$ beginning at $\tau_{0}-\delta$ for a duration of $\delta$ at which point it stays on for a detection time of $\Delta\tau\equiv\tau-\tau_{0}.$ Then, at time $\tau$, the function
$h_{2}$ begins to smoothly switch the detector off, again for a duration
of $\delta$, so that the detector returns to the off position at
$\tau+\delta.$ For a fixed $\Delta\tau$, the sharp switching limit is given 
by $\delta\to0$ which we pursue as a means of reducing unwanted transient effects associated with turning on the detector. These transient effects are particularly pronounced for trajectories with necessarily short detection times such as a detector plunging radially into a black hole, Ref.~\cite{Conroy13}.
\begin{figure}[htb]
	\centering
	\label{fig:onoff}
	\includegraphics[width=\linewidth]{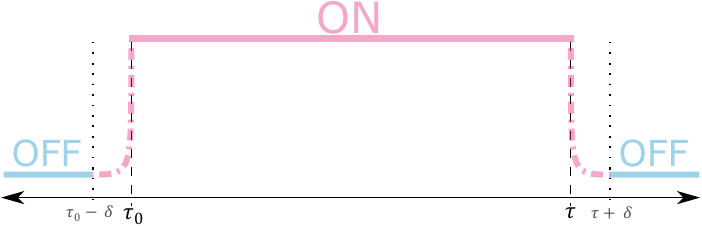}
	\caption{Schematic illustration of the on/off switching behaviour of the function $\chi(u)$ according to the definition and constraints given by~\eqref{switchdef} and~\eqref{hprops}. The detector begins in an `off' position, before being switched on smoothly, beginning at time $\tau_0-\delta$ for a duration of $\delta$. The dashed pink line illustrates this smooth switching behaviour, which tends towards a sharp switching limit as $\delta$ approaches zero. After the detector is fully switched on at time $\tau_0$, it remains in an `on' position, until time $\tau$, when the detector is smoothly turned off over a duration of $\delta$.}
\end{figure}

Following the recipe of Ref.~\cite{Satz2007}, we find the response function at the sharp-switching limit to be given by
\begin{align}
	\label{pert0}
	{\cal F}(\omega)&=\frac{1}{2\pi^{2}}\int_{\tau_{0}}^{\tau}du\int_{0}^{u-\tau_{0}}ds\left(\frac{\cos\omega s}{a(u)a(u-s)(\Delta x)^{2}}+\frac{1}{s^{2}}\right)
	\nn\\&+\frac{1}{2\pi^{2}}\ln\left(\frac{\Delta\tau}{\delta}\right)-\frac{\omega}{4\pi}\Delta\tau+C_{1}+{\cal O}\left(\frac{\delta}{\Delta\tau}\right)
	,\end{align}
for some constant $C_1$. Here we have perturbed around the small parameter $\delta/\Delta\tau$ where $\delta$ is the interval of smooth-switching.\footnote{The limit $\delta\to0$ is taken assuming that the detection time $\Delta\tau$ is finite. It is worth noting, however, that the perturbation \eqref{pert0} is also valid for finite $\delta$ and large detection time $\Delta\tau$.} For instantaneous switching we require $\delta\to0$ and here we see an explicit logarithmic divergence in the response function at this limit. However, all is not lost, as we can take the derivative with respect to $\tau$ to arrive at the \emph{transition rate}
\begin{align}
	{\cal \dot{F}}_{\tau}(\omega)&=\frac{1}{2\pi^{2}}\int_{0}^{\Delta\tau}ds\left(\frac{\cos\omega s}{a(u)a(u-s)(\Delta x)^{2}}+\frac{1}{s^{2}}\right)
	\nn\\&+\frac{1}{2\pi^{2}\Delta\tau}-\frac{\omega}{4\pi}+{\cal O}\left(\frac{\delta}{(\Delta\tau)^{2}}\right),
\end{align}
which is indeed well-defined for both instantaneous switching and infinite detection time. Thus, the transition rate we will analyse for a detector travelling through an FLRW spacetime is given by 
\begin{align}
	\label{transitionsharpAPP}
	{\cal \dot{F}}_{\tau}(\omega)=\frac{1}{2\pi^{2}}\int_{0}^{\Delta\tau}ds\left(\frac{\cos\omega s}{\sigma^2(\tau,s)}+\frac{1}{s^{2}}\right)+\frac{1}{2\pi^{2}\Delta\tau}-\frac{\omega}{4\pi},
\end{align}
where we have written the two-point Wightman function  in terms of the proper time parameters $u=\tau$ and $s$, with the latter encoding the history of the detector, while the geodesic distance $\sigma^2(\tau,s)\equiv a(\tau)a(\tau-s)(\Delta x)^{2}$ contains the contribution of the spacetime and trajectory. 
	\section{Cosmological Horizons}
\label{sec:Horizons}
For our purposes here, we work within a geometrically-flat FLRW universe with line element
\begin{equation}
	\label{FLRWapp}
	ds^2 =-dt^2 +a^2 (t)\left(dr^2+r^2d\Omega^2\right),
\end{equation}
and define the \emph{expansion} via the divergence
\begin{equation}
	\theta\equiv\nabla_\mu n^\mu=\frac{1}{\sqrt{-g}}\partial_\mu\left(\sqrt{-g}n^\mu\right),
\end{equation}
for some null ray $n^\mu$. By appealing to the geodesic equation for null tangent vectors, we find the ingoing null ray $n^\mu$, and its associated outgoing ray $l^\mu$, to be given by
\begin{equation}
	\label{inoutrays}
	n^\mu=\left(\frac{1}{a},-\frac{1}{a^2},0,0\right),\quad 	l^\mu= \left(\frac{1}{a}, \frac{1}{a^2},0,0\right),
\end{equation} 
where we have used the fact that the determinant of metric~\eqref{FLRWapp} is given by $g=\det g_{\mu\nu}=-a^6 r^4 \sin^2 \phi$. The ingoing and outgoing expansions are then
\begin{equation}
	\label{expin}
	\theta_{IN}=\frac{2}{a}\left(H-\frac{1}{\tilde r}\right),\quad 	\theta_{OUT}=\frac{2}{a}\left(H+\frac{1}{\tilde r}\right),
\end{equation}
where $\tilde r\equiv ar$ is the areal radius.

An apparent horizon is formed when the ingoing expansion vanishes and the outgoing expansion remains positive. In this context, we call such a surface the \emph{cosmological apparent horizon}\footnote{This is not to be confused with the particle horizon, which is the maximum distance a particle can travel along a geodesic in proper conformal time, i.e. $r_{PH}=\int^t_0 \frac{dt^\prime}{a(t^\prime)}$, c.f. Eq.~\eqref{FLRWconfapp} for relation to conformal time $\eta$.} and it forms the boundary of the minimally anti-trapped surface, i.e. the anti-trapped surface of minimal size. As such, setting the ingoing expansion~\eqref{expin} to zero yields an apparent horizon with areal radius
\begin{equation}
	\label{rAH}
	\tilde r_{AH}=H^{-1}\quad\mbox{where}\quad \tilde r_{AH}\equiv ar_{AH}.
\end{equation}

From the above, we can deduce that when $\tilde r>\tilde r_{AH}$ both the ingoing and outgoing expansions are positive, i.e. $\theta_{IN,OUT}>0$. The surfaces described by the expansion in this region are called \emph{anti-trapped}, while surfaces in the region $0\leq \tilde r < \tilde r_{AH}$, with $\theta_{OUT}>0$ and $\theta_{IN}<0$, are called \emph{normal surfaces}. In simple terms, outgoing geodesics in the normal region trace out a surface of larger area while ingoing geodesics trace out a shrinking surface with this being the familiar behaviour in flat space. We visualise the cosmological apparent horizon which forms the border between the normal and anti-trapped regions by considering an observer centred on a sphere, which we have positioned at $r=0$. Events beyond the sphere are causally disconnected from our observer, meaning information is obscured, see Ref. \cite{Faraoni:2015ula} for a more detailed discussion on cosmological horizons. 

\paragraph*{A note on terminology.} In place of the terminology used above, apparent horizons are often categorised as either future or past. Thus an apparent horizon formed by vanishing outgoing expansion is a \emph{future apparent horizon}, while an apparent horizon characterised by vanishing ingoing expansion, such as the cosmological apparent horizon described above, would be a \emph{past apparent horizon}. 

While it may seem that throughout the text we have used the terms `apparent horizon' and `trapped surface' interchangeably, these are in fact distinct. While both trapping surfaces and apparent horizons are defined by vanishing expansion, the definition of a trapping surface comes with an additional condition which distinguishes between inner and outer trapped surfaces, see Ref.~\cite{Faraoni:2015ula} for further details. 

For example, the \emph{future-outer trapped surface} of a black hole is formed at the locus whereby outgoing expansion vanishes (and the ingoing expansion remains negative), while a \emph{past-outer trapping surface} would be present in a white hole. To complete the picture, a \emph{past-inner trapping surface} of an expanding cosmology is formed at the locus whereby the ingoing expansion vanishes (and the outgoing expansion remains positive), while one could also picture a \emph{future-inner trapping surface} in a contracting cosmology, see Ref.~\cite{Helou:2015yqa} for further details. 

In this letter we consider only expanding cosmologies and so dispense with the inner/outer distinction and refer to the past-inner trapping surface as the cosmological apparent horizon. 
\\
\section{Fodor surface gravity in cosmology}
\label{sec:Fodorapp}
By defining some scalar quantity $f\equiv f(t)$ via $a(t)\equiv e^{f}$, we may write the conformally-flat FLRW metric given in Eq.~\eqref{FLRW} metric as
\begin{equation}
	ds^{2}=-dt^{2}+e^{2f(t)}dr^{2}+\tilde{r}^{2}d\Omega^{2},
\end{equation}
where $\tilde r=ar$ is the areal radius. As $r=\tilde{r}e^{-f(t)}$, we find $ dr=d\tilde{r}e^{-f}-\dot{f}\tilde{r}e^{-f}dt$ and so
\begin{equation}
	ds^{2}=\left(-1+\dot{f}^{2}\tilde{r}^{2}\right)dt^{2}-2\dot{f}\tilde{r}dtd\tilde{r}+d\tilde{r}^{2}+\tilde{r}^{2}d\Omega^{2}.
\end{equation}
Next, as $a=e^{f}$ implies $f=\ln a$ then $\dot{f}=H$ so that the FLRW metric in Painlev\'e-Gullstrand (PG) coordinates is given by
\begin{equation}
	\label{FLRWPG}
	ds^{2}=\left(-1+H^{2}\tilde{r}^{2}\right)dt^{2}-2H\tilde{r}dtd\tilde{r}+d\tilde{r}^{2}+\tilde{r}^{2}d\Omega^{2}.
\end{equation}
In these coordinates, the null ingoing and outgoing tangent vectors given in Eq.~\eqref{inoutrays} become
\begin{equation}
	\label{PGnullrays0}
	n^\mu=\frac{1}{a}\left(1,-1+H\tilde r,0,0\right),\quad 	l^\mu=\frac{1}{a}\left(1,1+H\tilde r,0,0\right).
\end{equation}
These geodesics give a (dynamical) cross normalisation of $n^\mu l_\mu=-2/a^2$. The ingoing and outgoing expansions in this case are
\begin{equation}
	\label{inoutexpapp}
	\theta_n=-\frac{2}{\tilde r}(1-H\tilde r)\quad\mbox{and}\quad \theta_l=\frac{2}{\tilde r}(1+H\tilde r),
\end{equation}
from which we observe that the ingoing expansion vanishes when $\tilde r=1/H$ while the outgoing expansion vanishes when $\tilde r=-1/H$. This allows us to define the surfaces
\begin{equation}
	\tilde	r_{AH}\equiv H^{-1}\quad\mbox{and}\quad \tilde r_{OT}\equiv -H^{-1},
		\label{horizons}
\end{equation}
where $\tilde r_{AH}$ is the cosmological apparent horizon, as before, and we have supposed the existence of an outer-trapped surface $\tilde r_{OT}$. 

To keep within the Fodor approach, however, we must choose alternative null tangent vectors to Eq.~\eqref{PGnullrays0} which give a normalisation that is a constant. As in Ref.~\cite{Nielsen:2005af}, we require a normalisation of $n^\mu l_\mu=-2$ and, as such, we  choose ingoing and outgoing null tangent vectors of the form  
\begin{equation}
	\label{PGnullrays1}
	\bar n^\mu=\left(1,-1+H\tilde r,0,0\right),\quad 	\bar l^\mu=\left(1,1+H\tilde r,0,0\right),
\end{equation}
which give the required normalisation. 

\acknowledgements{A.C. is supported by the Irish Research Council Postdoctoral Fellowship GOIPD/2019/536 and the Czech Science Foundation grant No 22-14791S. I would like to thank Dr Peter Taylor for many fruitful discussions and the referee for their helpful comments.} 
\bibliographystyle{apsrev4-2}
	\bibliography{allcitations}
\end{document}